\documentclass[aps,prl,showpacs,preprintnumbers,superscriptaddress,nofootinbib,floatfix,10pt,twocolumn]{revtex4-2}

\usepackage{amsfonts,amssymb,stmaryrd,latexsym,amsmath,braket}
\usepackage{graphicx,subfigure}
\usepackage{comment}
\usepackage{times}
\usepackage{slashed}
\usepackage{bm}
\usepackage{color}
\usepackage[colorlinks=true,linkcolor=blue,anchorcolor=red,urlcolor=blue,citecolor=red]{hyperref}

\newcommand{\beginsupplement}{%
        \setcounter{table}{0}
        \renewcommand{\thetable}{S\arabic{table}}%
        \setcounter{figure}{0}
        \renewcommand{\thefigure}{S\arabic{figure}}%
        \setcounter{equation}{0}
        \renewcommand{\theequation}{S\arabic{equation}}%
     }

\makeatletter
\let\saved@includegraphics\includegraphics
\AtBeginDocument{\let\includegraphics\saved@includegraphics}
\makeatother

\begin{document}

\title{Multi-neutron correlations in light nuclei via \emph{ab initio} lattice simulations}
\author{Shuang Zhang}\email{shu.zhang@fz-juelich.de}
\affiliation{Institute for Advanced Simulation (IAS-4), Forschungszentrum J\"{u}lich, D-52425 J\"{u}lich, Germany}

\author{Serdar Elhatisari}\email{selhatisari@gmail.com}
\affiliation{Interdisciplinary Research Center for Advanced Quantum Computing (IRC-AQC), King Fahd University of Petroleum and Minerals (KFUPM), 31261 Dhahran, Saudi Arabia}
\affiliation{Faculty of Natural Sciences and Engineering, Gaziantep Islam Science and Technology University, Gaziantep 27010, Turkey}

\author{Ulf-G. Mei{\ss}ner}\email{meissner@hiskp.uni-bonn.de}
\affiliation{Helmholtz-Institut f\"{u}r Strahlen- und Kernphysik and Bethe Center for Theoretical Physics, Universit\"{a}t Bonn, D-53115 Bonn, Germany}
\affiliation{Institute for Advanced Simulation  (IAS-4), Forschungszentrum J\"{u}lich, D-52425 J\"{u}lich, Germany}
\affiliation{Peng Huanwu Collaborative Center for Research and Education, International Institute for Interdisciplinary and Frontiers,
Beihang University, Beihang University, Beijing 100191, China}

\begin{abstract}

The quest to understand multi-neutron systems has a long history, and recent experimental efforts aim to probe candidate four-neutron configurations in neutron-rich light nuclei such as ${}^8$He and ${}^7$H via quasi-free knockout reactions.
However, the ground-state energies of the hydrogen isotopes ${}^6$H and ${}^7$H are not yet well constrained, with substantial discrepancies across experimental analyses and theoretical predictions.
Using ab initio nuclear lattice effective field theory with an ensemble of 282 chiral two- and three-nucleon forces, we perform a Bayesian uncertainty-quantified analysis of the ground-state energies of ${}^6$H and ${}^7$H. 
The marginal posteriors suggest single-neutron separation energy $S_n({}^{7}\mathrm{H})=0.35^{+0.32}_{-0.32}$ MeV, which kinematically disfavors sequential decay via ${}^{6}\mathrm{H}+n$ and thereby makes multi-neutron emission channels comparatively more relevant.
Intrinsic densities indicate triton- and $\alpha$-like clusters in ${}^7$H and ${}^8$He, respectively. 
By computing two-body and reduced four-body correlation functions, we find that the valence neutrons in the surface region of these systems form compact dineutrons that predominantly organize into approximately symmetric dineutron-dineutron configurations, with only a small but non-negligible fraction assembling into more compact tetraneutron-like substructures.
In ${}^7$H, these components account for roughly 95\% and 5\% of the sampled four-neutron configurations, respectively, and ${}^8$He exhibits a similar hierarchy.
For these configurations, we also extract the corresponding spatial and angular correlation patterns among the nucleons. These results provide nuclear-structure insights into the debate surrounding four-neutron clusters and complement ongoing experimental searches for tetraneutron signatures in light nuclei.

\end{abstract}
\maketitle
\date{today}

\paragraph{}{\itshape Introduction:} 
The nature of multi-neutron systems, in particular candidate three-neutron ($3n$) and four-neutron ($4n$) clusters, has been the subject of long-standing interest and controversy for more than half a century~\cite{Marques:2021mqf}.
Experimentally, our knowledge of such systems relies predominantly on indirect kinematic reconstructions in breakup~\cite{Marques:2001wh}, double charge-exchange~\cite{Kisamori:2016jie}, and quasi-free knockout reactions~\cite{Duer:2022ehf}, which are often discussed within a picture invoking preformed four-neutron configurations in neutron-rich projectiles. A fully exclusive observation of a correlated multi-neutron cluster remains a major challenge and the focus of ongoing efforts~\cite{Huang:2021lch}.
Such systems provide a unique window into many-body correlations in neutron-rich nuclei and dilute nuclear matter, and their direct exploration would clarify both their existence and their internal multi-neutron correlations, thereby linking multi-neutron correlations to more general many-body correlations in nuclei and nuclear matter.

More generally, correlations among nucleons in atomic nuclei are central to our understanding of nuclear structure and reactions, as well as nuclear astrophysics~\cite{Oertel:2016bki,Ye:2024slx}. Over the past decades, major experimental progress has been made in probing these correlations, including the detailed characterization of short-range correlated (SRCs) $np$ pairs~\cite{Hen:2016kwk,Arrington:2022sov}, long-range pairing correlations, manifest in halo nuclei and in nuclear superfluidity~\cite{Tanihata:1985psr,Nakamura:2006zz,Tanihata:2013jwa,Kikuchi:2016ofi,R3B:2018lfr,Kubota:2020pxo,Corsi:2023dek,Monteagudo:2024obw,Hammen:2018iny,Miller2019,Geldhof:2022,SAMURAI21-NeuLAND:2024kah}, observations of correlated two-nucleon emission from exotic nuclei~\cite{Blank:2007zz,Pfutzner:2011ju,Spyrou:2012zz,Kohley:2013kxt,Pfutzner:2023tvr}, and signatures of multi-nucleon correlations in finite nuclei~\cite{Freer:2017gip,Tanaka:2021oll}.
At the same time, modern radioactive ion beam facilities have produced multi-proton~\cite{Pfutzner:2023tvr} and multi-neutron~\cite{Kondo:2023F,Bezbakh:2019dvh,Muzalevskii:2020svp,A1:2025mjf} emitters, yet direct and exclusive observations of correlated multi-neutron emission remain severely limited by detector efficiency and acceptance, posing a major experimental challenge for future experiments and highlighting the need for complementary theoretical approaches.

On the theoretical side, two-body correlations have been studied extensively, including their spatial structure~\cite{Grigorenko:2003mc,Hagino:2005we,Hagino:2006ib,Papadimitriou:2011jx,Grigorenko:2013uqa,Kruppa:2013ala,Romero-Redondo:2016qmc,Wang:2017qij,Ma:2020roi} and dynamical properties~\cite{Bertulani:2008gq,Delion:2013qwa,Oishi:2014dna,Oishi:2016avj,Wang:2021atf}. 
Multi-nucleon correlations beyond the two-body level have primarily been investigated within Hartree–Fock–Bogoliubov approaches~\cite{Hagino:2008vm}, few-body models~\cite{Garrido:2025qfv}, antisymmetrized molecular dynamics (AMD) calculations with explicit cluster projection~\cite{Kanada-Enyo:2007iri,Aoyama:2009zz}, and microscopic cluster calculations based on the generator coordinate method~\cite{Itagaki:2008zz,Kobayashi:2013iwa,Yamaguchi:2023xsx,Nakagawa:2025zls}, all of which have yielded important insights into dineutron ($2n$) and $2n$–$2n$ correlations in neutron-rich nuclei.
In the SRCs regime, recent {\em ab initio} quantum Monte Carlo studies have probed three-nucleon correlations in the high-momentum components of nuclear wave functions~\cite{Weiss:2023laq}.
However, a direct calculation of four-body (and higher) correlation functions in nuclear systems remains highly nontrivial, because reduced density matrices beyond the two-body level suffer from a combinatorial explosion in dimensionality and memory requirements.
Further progress may thus be achieved with methods that can access $A$-body correlations without explicit construction of the full $A$-body density matrix. In this work, we exploit auxiliary-field Monte Carlo (AFMC) sampling to compute multi-neutron observables via Euclidean-time projection.

Motivated by quasi-free knockout experiments~\cite{JACOB:1966wce,Jacob:1973tu,Yoshida:2025ptaf119}, which have revealed evidence for preformed clusters in neutron-rich nuclei~\cite{Tanaka:2021oll} and have suggested that quasi-free knockout can access four-neutron configurations that closely resemble the intrinsic $4n$ subsystem of the projectile nucleus~\cite{Duer:2022ehf}, we focus on neutron-rich $^{7}$H and $^{8}$He. These cluster-core systems with four valence neutrons provide unique laboratories for studying emergent $4n$ correlations, and our analysis of their internal correlations offers complementary nuclear-structure insights into the long-standing experimental and theoretical controversy surrounding $3n$ and $4n$ systems (see Ref.~\cite{Marques:2021mqf} and references therein and see Refs.~\cite{Faestermann:2022meh,Duer:2022ehf,Lazauskas:2022mvq,RIBF-SHARAQ11:2024cvz,Mazur:2024byg} for recent developments).
Here, we present an {\em ab initio} investigation of the neutron-rich nuclei $^{7}$H and $^{8}$He using nuclear lattice effective field theory (NLEFT)~\cite{Lahde:2019npb}, with full $A$-body correlations. 
We identify compact surface dineutrons that organize into two representative four-neutron configurations and quantify the geometry and probability of these configurations from two- and four-body correlation functions.

\paragraph{}{\itshape Methods:} NLEFT provides a powerful framework to capture the full set of many-body correlations through AFMC simulations, but its application to high-fidelity chiral interactions has been hindered by severe sign problems~\cite{Lahde:2019npb,Lu:2018bat,Lu:2021tab}. 
With recent advancements via the wavefunction matching method, chiral forces up-to-and-including N$^3$LO can now be treated largely free from sign problems~\cite{Elhatisari:2022zrb}.
In this work, we employ a set of N$^3$LO chiral interactions that include both nucleon–nucleon (NN) and three-nucleon (3N) forces. These interactions are obtained through history matching~\cite{Vernon:2010Galaxy,Vernon:2014Galaxy,Hu:2021trw,Elhatisari:2022zrb} that systematically incorporates uncertainties from NN scattering data, the truncation of the chiral hierarchy, and the binding energies of selected nuclei.
The main results are derived from calculations using the best-fit interaction, while the corresponding uncertainties and correlations among observables are quantified using Bayesian posterior predictive distributions (PPD) constructed from an ensemble of 282 non-implausible chiral interactions, as detailed in Supplemental Material (SM)~\cite{supp}.

To probe observables associated with nuclear density, particularly those related to intrinsic structure and correlation patterns, we employ the pinhole method~\cite{Elhatisari:2017eno}. This approach enables sampling of the spatial coordinates, spins, and isospins of all nucleons according to the quantum amplitudes of the $A$-body density within the auxiliary-field framework. With these pinhole configurations, a wide range of observables of physical interest can be evaluated. Details of the sampling scheme and observable evaluation are described in SM~\cite{supp}. In the following, expectation values are denoted by $\langle \hat O \rangle \equiv \langle\Psi|\hat O|\Psi\rangle$.
In order to reveal the intrinsic structure of the nucleus, an appropriate alignment scheme for the intrinsic density  is often employed, as in Refs.~\cite{Wiringa:2000gb,Elhatisari:2017eno,Otsuka:2022bcf,Shen:2022bak,Shen:2024qzi}. Here, we identify candidate clusters (such as an $\alpha$ particle or a triton) by selecting the most probable grouping of nucleons that is consistent with the cluster spin-isospin quantum numbers and spatial compactness within each sampled pinhole configuration. Once the clusters are identified, we realign the density profile with respect to the cluster positions, thereby making the intrinsic spatial structure of the nucleus more apparent. Further technical details are provided in SM~\cite{supp}.

Following the identification of the intrinsic clusters, we extend our analysis to multi-neutron correlations. In contrast to compact clusters, these correlations exhibit more delocalized patterns. We begin with the two-body correlation function,
\begin{equation}
\begin{aligned}
\rho_{2}(r,r',\theta)
= \sum_{i<j}
  \Big\langle
    \delta(r_i - r)\,
    \delta(r_j - r')\,
    \delta(\theta_{ij}-\theta)
  \Big\rangle ,\\
\rho_{2}^{(\varphi,\theta)}(\varphi,\theta)
= \sum_{i<j}
  \Big\langle
    \delta(\varphi_{ij}-\varphi)\,
    \delta(\theta_{ij}-\theta)
  \Big\rangle,
\end{aligned}
\end{equation}
where $r_i, r_j$ denote the distances of the two nucleons $i,j$ from the center of mass (c.m.), $\theta_{ij}$ is the opening angle between them, and $\varphi_{ij}$ represents the angle between the relative position vector of the two nucleons and the bisector of their opening angle.
The first form, similar to those used in previous studies~\cite{Bertsch:1991zz,Hagino:2005we,Papadimitriou:2011jx}, characterizes the spatial distribution of two nucleons in terms of their relative distances from the c.m. and opening angle. The appropriate Jacobian is included when transforming from the pinhole coordinates. The second form, although it lacks an explicit radial-distance dependence, highlights the radial asymmetry between the two nucleons. When $\varphi_{ij}=90^\circ$, the configuration is approximately symmetric, corresponding to two nucleons at similar distances from the c.m. This quantity is further utilized in the subsequent analysis of multi-neutron correlations.

In a four-body subsystem, correlations span many degrees of freedom, including individual coordinates and all pairwise opening angles, making it difficult to extract clear structural information directly. Building on the two-body correlations and symmetry considerations, we construct reduced projections that capture the essential features of the underlying multi-nucleon correlations. In particular, we define a reduced four-body correlation function,
\begin{equation}\label{eq:4ncor}
\begin{aligned}
\rho_4(\theta_1,\varphi_1,\theta_2,\varphi_2;\Theta,\zeta) =& 
\sum_{i<j,\,k<l}^{(ij),(kl)}
  \Big\langle
    \delta(\theta_{ij} - \theta_1)\,
    \delta(\varphi_{ij} - \varphi_1)\\
&\times \delta(\theta_{kl} - \theta_2)\,
       \delta(\varphi_{kl} - \varphi_2)\\
&\times \delta(\Theta_{ij,kl} - \Theta)\,
       \delta(\zeta_{ij,kl}-\zeta)
  \Big\rangle,
\end{aligned}
\end{equation}
where $i,j,k,l$ label the nucleons, the sum runs over pairs of disjoint nucleon pairs $(ij)$ and $(kl)$, and the angles $\theta_{ij}, \varphi_{ij}$ and $\theta_{kl}, \varphi_{kl}$ for each pair are defined in the same way as in the two-body case. The angle $\Theta_{ij,kl}$ is the opening angle between the c.m. vectors of the two pairs, while $\zeta_{ij,kl}$ denotes the torsion angle between the relative position vectors of nucleons in each pair.
This definition projects the four-body density onto configurations in which two disjoint nucleon pairs exhibit a specified angular geometry, as schematically illustrated in Fig.~S1 of SM~\cite{supp}.

In practice, we perform AFMC simulations on a periodic cubic lattice with side length of $L = 12$ and spatial lattice spacing of $a \simeq 1.32$~fm, corresponding to a momentum cutoff of $\Lambda = \pi/a \simeq 471$~MeV. The temporal lattice spacing is set to $a_t = 0.001$~MeV$^{-1}$ for the Euclidean-time evolution.  
The initial states are chosen as the product of the single-particle harmonic oscillator (HO) wavefunctions. For the hydrogen isotopes, the HO frequency of the valence neutrons is reduced to generate more diffuse wavefunctions and to accelerate convergence. Calculated energies are extrapolated to the infinite Euclidean-time limit using the formalism outlined in SM~\cite{supp}. 
Finite-volume effects were explicitly assessed at the present statistical precision and are found to be subdominant compared with statistical uncertainties~\cite{supp}. We therefore do not perform a volume extrapolation.
In the pinhole calculation, the point-nucleon coordinates are convolved with a Gaussian~\cite{Elhatisari:2017eno} of width $0.84$~fm to account for the finite size of the nucleons~\cite{Lin:2021xrc}. Each pinhole configuration undergoes 3000 Gaussian smearing iterations to enhance the smoothness and reduce lattice artifacts in the density profiles. Unless stated otherwise, density-related observables are evaluated at a Euclidean projection time of $\tau = 0.35$~MeV$^{-1}$, ensuring a balance between convergence and statistical precision, following Refs.~\cite{Elhatisari:2017eno,Shen:2022bak,Shen:2024qzi,Zhang:2024wfd}.

\paragraph{\itshape Results:}

\begin{figure}[t]
\centering
      \includegraphics[width=0.48\textwidth]{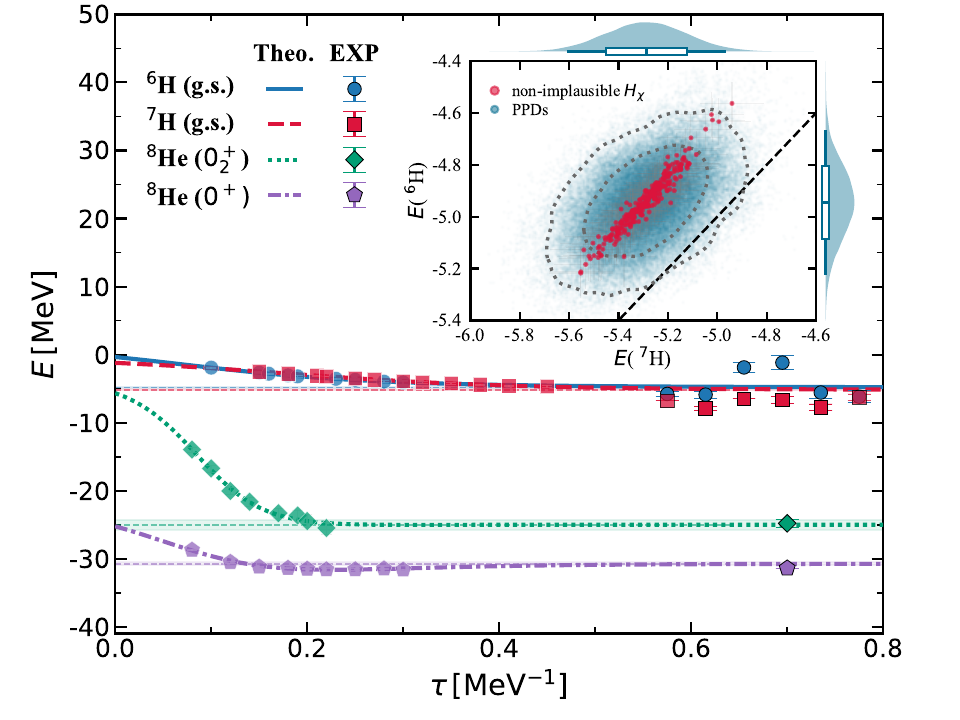}\\
  \caption{Calculated and experimental g.s. energies of $^{6,7}$H, and the 0$^{+}$ and 0$_2$$^{+}$ state energies of $^{8}$He.
  Colored markers overlaid on the corresponding curves indicate the calculated energies versus Euclidean time $\tau$, with statistical uncertainties. The curves represent the $\tau$-extrapolation fits, and horizontal dashed lines denote the extrapolated $\tau\!\to\!\infty$ energies with their statistical uncertainties. 
  Experimental data~\cite{Aleksandrov:1984H6,Belozyorov:1986H456,Gurov:2003pv,Caamano:2008zz,A1:2025mjf,Korsheninnikov:2000He7unique,Caamano:2007zz,Nikolskii:2010zz,Bezbakh:2019dvh,Caamano:2020,Muzalevskii:2020svp} are shown as symbols labeled ``EXP'' in the legend.
  \textbf{Inset:} The red circles represent the g.s. energies of $^{6,7}$H from non-implausible prior samples obtained via history matching \cite{Elhatisari:2022zrb}, including uncertainties. Posterior samples are shown as pale blue points, with dashed lines indicating the 1$\sigma$ and 2$\sigma$ credible intervals. The dashed black line denotes the condition $E(^{6}\mathrm{H}) = E(^{7}\mathrm{H})$. Marginal distributions are shown along each axis, with medians indicated by ticks and thick and thin bars representing the 1$\sigma$ and 2$\sigma$ credible intervals, respectively. }
  \label{fig:E}
\end{figure}

The calculated energies of $^{6,7}$H and $^{8}$He are shown in Fig.~\ref{fig:E} and compared with experimental data.
For $^{8}$He, multichannel calculations projected onto the $A_1^+$ {\em irrep} of the cubic group~\cite{Johnson:1982yq,Lu:2014xfa} yield energies of $-30.73(19)$~MeV and $-24.99(72)$~MeV for the $0^+$ and $0_2^+$ states, respectively, consistent with the experimental values $-31.41(1)$~MeV~\cite{Tilley:2004zz} and $-24.74(6)$~MeV~\cite{Yang:2023hyq}.
In addition, the lattice simulations reveal a dineutron correlation in $^{8}$He that emerges and stabilizes at sufficiently long Euclidean times, suggesting configuration mixing beyond the initial pure $(0p_{3/2})^4$ filling.
A detailed discussion of this pattern is given below.
This aligns with reaction~\cite{Keeley:2007bc,Skaza:2007fus,Holl:2021bxg} and structure~\cite{Fossez:2018gae,Yamaguchi:2023xsx,Nakagawa:2025zls} studies that indicate configuration mixing and deformation in $^{8}$He, and with the newly observed $0_2^+$ cluster-like excitation, for which strong $2n$–$2n$ components have been discussed~\cite{Yang:2023hyq}.
Extending to hydrogen isotopes, we find a $J^\pi=1^+$ ground state (g.s.) in $^{6}$H, suggesting $\nu1s_{1/2}$ occupancy and a breaking of the $0p_{3/2}$ subshell closure, which is conducive to dineutron correlations~\cite{Papadimitriou:2011jx}.
Experimentally, the reported $^{6,7}$H g.s. energies and widths vary across reactions (see Table~S2 in SM~\cite{supp}), and two advanced theoretical approaches predict $^{6,7}$H g.s. energies that differ by about 7~MeV~\cite{Li:2021tyy,Hiyama:2022gzv}, underscoring the sensitivity to the interaction and many-body treatment.
Incorporating full $A$-body correlations, we perform a Bayesian analysis over 282 non-implausible chiral interactions to quantify uncertainties. The resulting g.s. energy medians and $1\sigma$ credible intervals for $^{6,7}$H are $-4.94^{+0.14}_{-0.14}$ MeV and $-5.29^{+0.16}_{-0.16}$ MeV, respectively, within the experimental ranges. A quantitative treatment of resonance widths requires a dedicated analysis and is beyond the present scope.
The marginal posterior for the one-neutron separation energy of $^{7}$H suggests a positive $S_n$ at the 1$\sigma$ level, $S_n = 0.35^{+0.32}_{-0.32}$~MeV, which kinematically disfavors sequential decay via $^{6}$H$+n$ and thus makes multi-neutron emission channels comparatively more relevant.
These findings, together with experimental and theoretical indications~\cite{Duer:2022ehf,Lazauskas:2022mvq} of dineutron clustering in $4n$ dynamics, underscore the emergence of enhanced multi-neutron correlations in $^{7}$H and $^{8}$He, from dineutron to possible tetraneutron-like four-body configurations, and set the stage for the correlation analysis that follows.

\begin{figure}[t]
  \centering
  \subfigure{
    \includegraphics[width=0.45\textwidth]{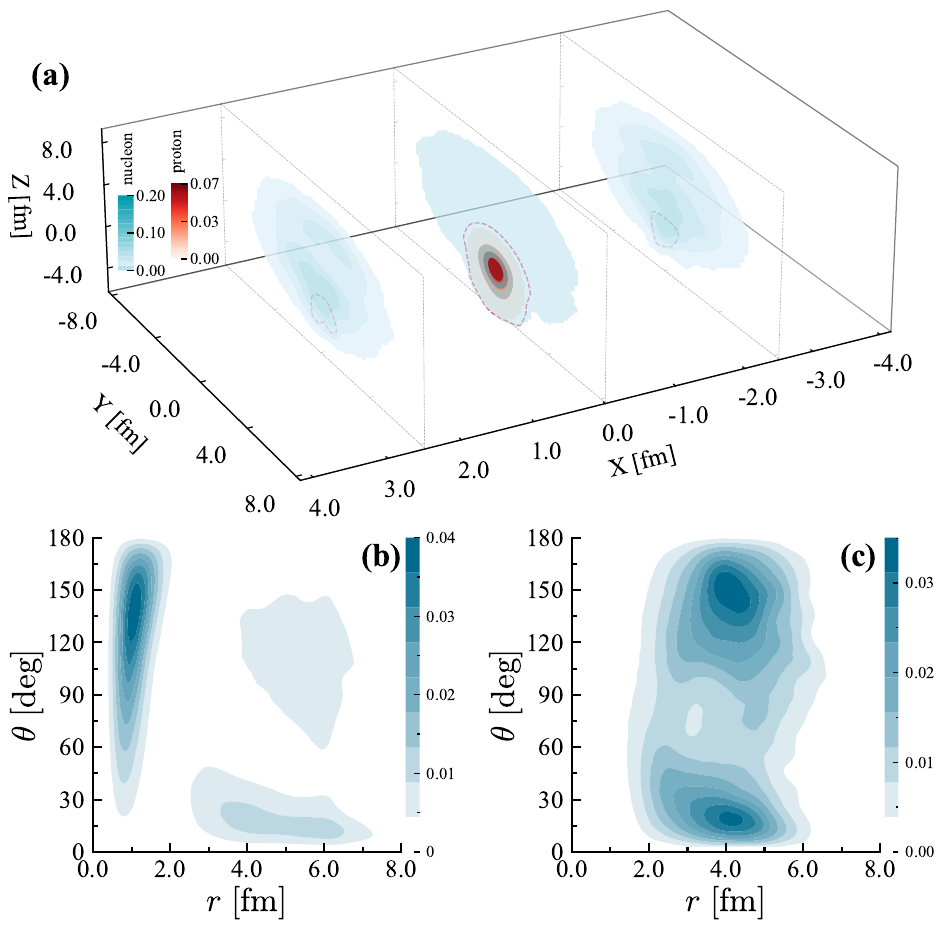}
  }
  \caption{%
    (a) Intrinsic nucleon (blue) and proton (red) densities of the $^7$H ground state, visualized as slices at $x=0$ and $x=\pm 2.5$~fm. The red dashed line marks the maximal spatial extent of the identified triton cluster. (b) Two-neutron correlation density of all neutrons in the $^7$H ground state, with $r$ defined relative to the c.m. of the core. (c) Two-neutron correlation density of the four valence neutrons, with $r$ defined with respect to the c.m. of the $^7$H.}
  \label{fig:D}
\end{figure}

Fig.~\ref{fig:D} presents the intrinsic nucleon densities and two-neutron ($nn$) correlations in $^{7}$H, taken as a representative case of multi-neutron dynamics near the dripline (see SM~\cite{supp}, Figs.~S2 for benchmarks in $^{6}$He and Figs.~S3 for a complementary analysis of $^{8}$He).
The intrinsic frame is defined with the triton c.m. located on the negative $z$ axis, so that the valence neutrons in this convention predominantly appear in the upper (positive-$z$) half of the panels.
The nucleon and proton density maps on the $x=0$ and $\pm2.5~\mathrm{fm}$ slices reveal a compact triton cluster surrounded by four diffuse valence neutrons. This clustered geometry is consistent with a core–plus–valence-neutron picture as commonly employed in cluster and many-body approaches.
The interpretation is further supported by Fig.~\ref{fig:D}(b), which displays $nn$ pair correlations for all six neutrons relative to the triton c.m. and reveals two distinct components.
The inner component is concentrated at short distances with only mild angular modulation, analogous to the $nn$ correlations in $^3$H~\cite{supp}. 
The outer component emerges at larger radii, where pronounced enhancements are observed at both small and large opening angles $\theta$ between the two neutrons.
These two distinct components indicate that our core–valence partition has only a limited impact on the subsequent correlation analysis and, in particular, that it allows us to analyze the correlations among the valence neutrons.
Motivated by this apparent separation, we extract the two-body correlations among the valence neutrons in the $^{7}$H c.m. frame, Fig.~\ref{fig:D}(c), as commonly used in experimental analyses, and resolve more detailed features of the valence-neutron angular correlations. 
The resulting bimodal angular pattern closely resembles that obtained in Fig.~\ref{fig:D}(b), indicating the robustness of the dineutron correlations, and shows pronounced maxima around $\theta\simeq 21^\circ$ and $151^\circ$ at $r\simeq 4$~fm.
At small $\theta$, the enhancement indicates the presence of compact dineutron-like pairs in the surface region. 
We further extract the typical $2n$ size for pairs with small $\theta$, as detailed in SM~\cite{supp} (Figs.~S3).
The comparable strength at large $\theta$ contrasts with systems such as $^{6}$He, where only a relatively weak large-$\theta$ component has been reported~\cite{Hagino:2005we,Papadimitriou:2011jx,Assie:2010zz}.
For $^{7}$H, each pinhole configuration contains several neutron pairs, so the two-body correlation function represents a statistical distribution over all pairs and thus provides a probe sensitive to the underlying multi-neutron geometry.
The enhanced large-$\theta$ component indicates that, besides compact surface dineutrons, neutron pairs also populate more extended, back-to-back geometries.
This pattern is qualitatively consistent with configurations in which the four valence neutrons form two dineutron-like clusters located on opposite sides of the core, in line with a more symmetric $2n$–$2n$ geometry inferred from experimental analyses of $^{8}$He~\cite{Mueller:2007dhq} and supported by microscopic calculations of $^{8}$He and $^{7}$H that exhibit $2n$–$2n$ clustering~\cite{Hagino:2008vm,Garrido:2025qfv,Kanada-Enyo:2007iri,Aoyama:2009zz,Itagaki:2008zz,Kobayashi:2013iwa,Yamaguchi:2023xsx,Nakagawa:2025zls}.
Such bimodal angular correlations are suggestive of $2n$–$2n$ clustering and may represent a microscopic mechanism underlying enhanced $4n$ correlations in $^{7}$H and $^{8}$He.

\begin{figure*}[ht]
\centering
      \includegraphics[width=0.95\textwidth]{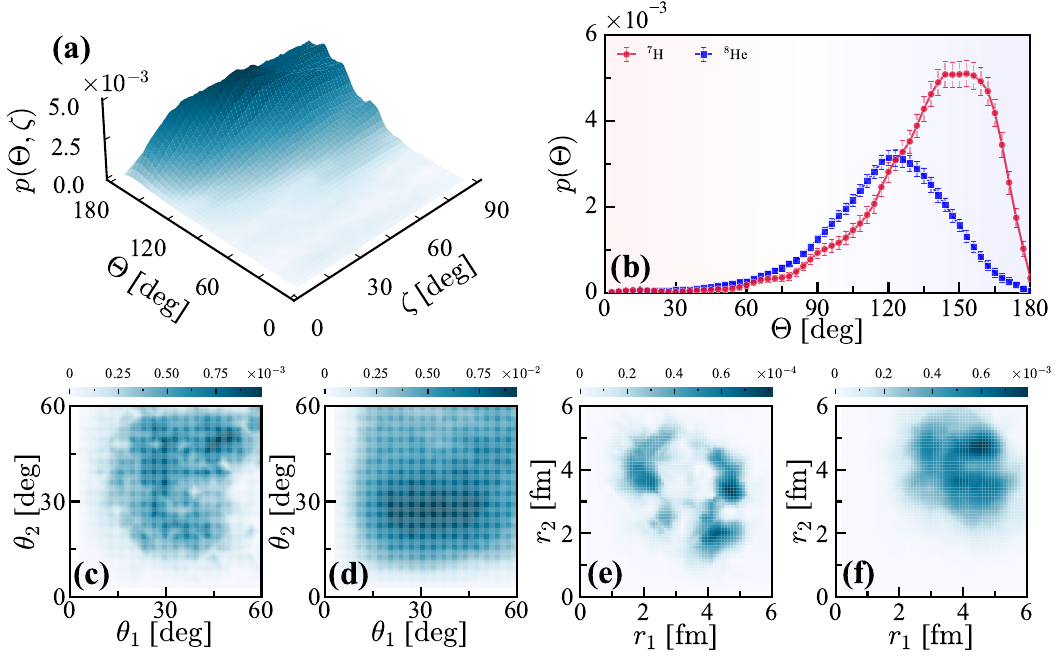}\\
  \caption{Four-neutron geometry and correlations.
  (a) Joint probability density $p(\Theta,\zeta)$ of the inter-pair opening angle $\Theta$ and torsion angle $\zeta$ in $^{7}$H.
(b) Comparison of the inter-dineutron opening-angle probability density $p(\Theta)$ in $^{7}$H and $^{8}$He under the same intra-angle selection, $\theta_{1,2}\le60^\circ$ and $|\varphi_{1,2}-90^\circ|\le10^\circ$, where we classify configurations with $\Theta<90^\circ$ as compact and $\Theta>90^\circ$ as extended for discussion.
(c-f) Conditional internal angular and radial correlations for configurations selected at $\Theta\approx69^\circ$ (compact) and $\Theta\approx141^\circ$ (extended). Here $\theta_{1,2}$ denote the intra-pair opening angles of the two selected $nn$ pairs, and $r_{1,2}$ denote the distances from the c.m. of each pair to the $^{7}$H c.m.
Angles are shown in degrees, while $p(\Theta,\zeta)$ and $p(\Theta)$ are reported per radian squared and per radian, respectively.}
\label{fig:4n}
\end{figure*}

The structure of $4n$ configurations has been explored analytically in schematic models~\cite{Zhukov:1994zz,Mei:2012aop} and also within cluster descriptions that treat the four neutrons as two correlated dineutron clusters~\cite{Hagino:2008vm,Garrido:2025qfv,Kanada-Enyo:2007iri,Aoyama:2009zz,Itagaki:2008zz,Kobayashi:2013iwa,Yamaguchi:2023xsx,Nakagawa:2025zls}. 
A systematic characterization of four-neutron geometries at the nucleonic level has been limited, and an explicit four-body treatment is required to access detailed spatial and angular information.
In practice, we evaluate Eq.~(\ref{eq:4ncor}) using pinhole samples with two spin-up and two spin-down valence neutrons, which constitute the bulk of the pinhole ensemble. Motivated by dineutron correlations seen in $^{7}$H and $^{8}$He, we characterize the four-neutron geometry using two opposite-spin pair coordinates sensitive to dineutron-like structure.
Fig.~\ref{fig:4n}(a) shows the marginal distribution of the inter-pair opening and torsion angles $(\Theta,\zeta)$ obtained from Eq.~(\ref{eq:4ncor}) after integrating over the intra-pair angles $(\theta_1,\varphi_1)$ and $(\theta_2,\varphi_2)$.
Guided by the $nn$ correlations of ${}^{7}$H discussed above, we identify dineutron-like configurations by requiring $\theta_{1,2}\le 60^\circ$ and $\Delta\varphi_{1,2}\le 10^\circ$, where $\Delta\varphi_{1,2}\equiv|\varphi_{1,2}-90^\circ|$ quantifies the deviation from radial symmetry within each pair. 
The resulting marginal distribution shapes depend only weakly on variations of the $(\theta,\Delta\varphi)$ cuts, see Figs.~S4 in SM~\cite{supp}.
The marginal probability density is strongly enhanced for opening angles $\Theta$ in the range $140^\circ$--$160^\circ$, and forms a broad maximum at $(\Theta,\zeta)\approx(141^\circ,48^\circ)$. 
With the same $\Theta$ range, the dihedral-angle distribution exhibits a peak around $30^\circ$--$40^\circ$ (see Fig.~S5 for details).
This pattern is suggestive of configurations containing two dineutron-like pairs on opposite sides of the system c.m., with nonzero torsion and dihedral angles indicating twisting and non-planarity.
While Ref.~\cite{Mei:2012aop} suggested symmetric and coplanar “great-circle’’ configurations in $^{8}$He based on a pure $(\nu0p_{3/2})^{4}$ shell-model occupation, our results show that such coplanar arrangements are strongly reduced once many-body correlations are treated explicitly.
These features suggest that the inter-$2n$ geometry itself is modified by many-body correlations.
Consequently, the four-neutron subsystem favors an extended, mildly twisted configuration that is approximately spatially symmetric about the system c.m.
In addition to this dominant extended configuration, tetraneutron-like geometrical components corresponding to the region of small opening angles, where the four neutrons reside on the same side of the system c.m., occur with lower probability.
As shown in Fig.~\ref{fig:4n}(b), this component accounts for about 5\% of the sampled configurations shown in panel~(b) and is concentrated at $\Theta$ in the range $60^\circ$--$90^\circ$.
Its associated torsion angle $\zeta$ and dihedral angle (see Fig.~S5 also lie predominantly in the range $60^\circ$--$90^\circ$, indicating a strongly noncoplanar and twisted four-neutron geometry compared to the symmetric $2n$--$2n$ configuration.
For comparison, the $\Theta$ distribution in $^{8}$He is shifted to smaller angles and more strongly suppressed at large $\Theta$, indicating a less prominent symmetric $2n$-$2n$ geometry than in $^{7}$H under the same selection.
This may be attributable to core-induced medium effects on the four-neutron configuration, in that the different proton background and core structure modify the environment experienced by the four neutrons.

To further elucidate these two geometries, Figs.~\ref{fig:4n}(c–f) display the internal angular and radial correlations for configurations corresponding to the extended, approximately symmetric $2n$–$2n$ geometry at $\Theta\approx141^\circ$ and the compact tetraneutron-like geometry at $\Theta\approx69^\circ$ (see Figs.~S6 in SM~\cite{supp} for additional details).
For the extended $2n$–$2n$ geometry, Figs.~\ref{fig:4n}(d,f), each dineutron pair retains a small internal opening angle $\theta$, with its distribution peaked near $30^\circ$, and the two pairs exhibit nearly symmetric distributions of the distances $r_{1,2}$ (defined as the distances from each pair c.m. to the system c.m.) that extend to about 4–5~fm.
In contrast, the tetraneutron-like geometry, Figs.~\ref{fig:4n}(c,e), shows enhanced probability density near characteristic radial-distance pairs $(2,4)$~fm and $(3,5)$~fm from the system c.m., indicating two radially asymmetric $nn$ pairs located on the same side of the system c.m.
Their internal opening-angle distributions are shifted to larger values, with a pronounced maximum near $\theta\approx60^\circ$ and lying predominantly in the range $30^\circ$-$60^\circ$. These configurations are also accompanied by large torsion and dihedral angles between $60^\circ$ and $90^\circ$ (see Fig.~S5). 
Together, these features lead to a non-coplanar, interlaced geometry.
This geometry represents a compact, strongly correlated four-neutron configuration, in which the distinction between the two $2n$ clusters becomes less pronounced, suggesting enhanced four-body correlations among the four neutrons.
It may be viewed as a structural precursor to a possible tetraneutron configuration.
Overall, these results highlight two representative geometries of the correlated four-neutron subsystem in $^{7}$H: an approximately spatially symmetric, extended $2n$-$2n$ configuration with large inter-pair separation and finite torsion, and a compact, twisted tetraneutron-like configuration characterized by enhanced four-body correlations. The extended $2n$-$2n$ geometry dominates the marginal $\Theta$ distribution at the $\sim95\%$ level, while the compact component contributes the remaining $\sim5\%$, with the relative weights only weakly sensitive to reasonable variations of the $(\theta,\Delta\varphi)$ selection (see Figs.~S4). The strong dominance of the extended $2n$–$2n$ geometry in the neutron-rich light nuclei studied here implies that experimental searches for tetraneutron-like configurations are likely to receive sizable contributions from symmetric $2n$--$2n$ modes, complicating the isolation of a genuinely compact, tetraneutron-like $4n$ signal.

\paragraph{}{\itshape Summary:} 
We perform {\em ab initio} NLEFT calculations of the neutron-rich nuclei $^{6,7}$H and $^{8}$He, including full $A$-body correlations and quantifying uncertainties with a Bayesian analysis over an ensemble of chiral interactions.
We predict the ground-state energies of the extreme hydrogen isotopes $^{6,7}$H and obtain the energies of the benchmark nucleus $^{8}$He consistent with available data, while simultaneously uncovering their internal multi-neutron structure.
Intrinsic densities and two-neutron correlation functions indicate a compact triton (alpha) cluster in $^{7}$H ($^{8}$He) surrounded by four valence neutrons that form compact dineutron pairs.
Building on this, we evaluate four-body correlations directly on lattice Monte Carlo configurations and construct a reduced four-body correlation function to probe the internal geometry of the four-neutron subsystem.
We extract the corresponding spatial and angular correlation patterns between each nucleon in two representative geometries.
We find that the $^{7}$H ground state is dominated (about 95\%) by an approximately spatially symmetric, extended $2n$–$2n$ geometry with large inter-pair separation and finite torsion, while a compact, twisted tetraneutron-like configuration with pronounced four-body correlations accounts for the remaining 5\%.
These results provide a microscopic picture of emergent multi-neutron correlations in $^{7}$H and $^{8}$He and 
suggest that experimental searches for tetraneutron-like signatures may receive non-negligible contributions from symmetric $2n$-$2n$ modes, which can complicate the isolation of a genuinely compact $4n$ signal.


\begin{acknowledgments}
We are grateful for discussions with Y. Z. Ma, S. H. Shen, Z. X. Ren, F. Hildenbrand, T. L\"ahde, A. Sarkar, and other members of the NLEFT collaboration. S. Zhang thanks the discussion with S. W. Huang, S. M. Wang, B. S. Hu, J. G. Li, and X. X. Sun.
This work is part of the EXOTIC grant and was supported in part by the European Research Council (ERC) under the European Union's Horizon 2020 research and innovation programme (grant agreement No. 101018170).
The work of UGM was also supported  by the CAS President's International
Fellowship Initiative (PIFI) (Grant No.~2025PD0022).
The work of SE is supported in part by the Scientific and Technological Research Council of Turkey (TUBITAK project no. 123F464).
The authors gratefully acknowledge the Gauss Centre for Supercomputing e.V. (www.gauss-centre.eu)
for funding this project by providing computing time on the GCS Supercomputer JUWELS
at J\"ulich Supercomputing Centre (JSC).  Furthermore, the authors gratefully acknowledge the computing time provided on the high-performance computer
HoreKa by the National High-Performance Computing Center at KIT (NHR@KIT). This center is
jointly supported by the Federal Ministry of Education and Research and the Ministry of Science,
Research and the Arts of Baden-Württemberg, as part of the National High-Performance Computing
(NHR) joint funding program (https://www.nhr-verein.de/en/our-partners). HoreKa is partly funded
by the German Research Foundation (DFG).
\end{acknowledgments}

\clearpage

\beginsupplement
\onecolumngrid
\section{Supplemental Material}

\subsection{Uncertainty quantification of chiral forces}

We start from 282 non-implausible N$^3$LO chiral interactions on the lattice~\cite{Elhatisari:2022zrb_sm}, obtained via history matching~\cite{Vernon:2010Galaxy_sm,Vernon:2014Galaxy_sm,Hu:2021trw_sm,Elhatisari:2022zrb_sm} that explicitly account for uncertainties in NN scattering data, the truncation of the chiral hierarchy, and binding energies of selected nuclei.
To construct the posterior distribution informed by experimental data, the first step typically involves incorporating additional observables associated with the relevant structural features, thereby conditionally refining the physical credibility of the non-implausible interaction set.
The posterior is subsequently obtained within a Bayesian framework that combines a Gaussian likelihood with a prior over the chiral interactions,
\begin{equation}
P(\boldsymbol{\lambda} \mid \mathcal{O}_\text{exp}) \propto 
P(\mathcal{O}_\text{exp} \mid \boldsymbol{\lambda})\, P(\boldsymbol{\lambda})\propto 
\exp\!\left[
-\tfrac{1}{2}
\frac{
\left(\mathcal{O}_\text{cal}(\boldsymbol{\lambda}) - \mathcal{O}_\text{exp}\right)^{2}
}{
\sigma_\text{cal}(\boldsymbol{\lambda})^{2} + \sigma_\text{exp}^{2}
}
\right]
 \exp\left[-\tfrac{1}{2}
\chi^2(\boldsymbol{\lambda}) 
\right]
\end{equation}
where $\boldsymbol{\lambda}$ denotes a specific chiral interaction from the non-implausible ensemble, $\mathcal{O}_\text{calc}(\boldsymbol{\lambda})$ and $\mathcal{O}_\text{exp}$ are the calculated and experimental observables, 
and $\sigma_\text{calc}(\boldsymbol{\lambda})$ and $\sigma_\text{exp}$ represent their theoretical and experimental uncertainties, respectively.
The prior $P(\boldsymbol{\lambda})$, derived from the $\chi^2$ values of the history-matching fits~\cite{Elhatisari:2022zrb_sm}, reflects the relative credibility of each interaction in reproducing the calibration data,  while the likelihood $P(\mathcal{O}_\text{exp} \mid \boldsymbol{\lambda})$ quantifies the agreement between the predictions of each interaction and the selected observables.

In this work, because the ${}^3$H binding energy relevant to the hydrogen isotopes was already used in the history-matching procedure that constrained the chiral interactions, the likelihood is approximated as uniform.
For each chiral interaction, we calculate the ground-state energies of ${}^{6,7}$H and extract their extrapolated values with statistical uncertainties.
The total uncertainty originates from several sources, including the many-body method approximation (first-order perturbative treatment in the wavefunction matching), finite-volume extrapolation, Euclidean-time projection, and statistical noise. The systematic uncertainty associated with the many-body method approximation is reduced through variational Hamiltonian optimization, resulting in a residual error smaller than the statistical one~\cite{Elhatisari:2022zrb_sm}.
As demonstrated below, finite-volume effects also lie within the statistical uncertainties. Therefore, we adopt the statistical uncertainty as the representative error throughout this work.
Based on the posterior distribution $P(\boldsymbol{\lambda} \mid \mathcal{O}_\text{exp})$, we generate 80000 random samples by sampling Gaussian fluctuations around the extrapolated values according to their standard deviations, yielding the final statistical distributions and correlations among the observables.

\subsection{Evaluation of observables with pinhole configurations}
In the pinhole method, we first introduce the normal-ordered $A$-body density operator $\rho_{i_1,j_1,\cdots,i_A,j_A}(\mathbf{n}_1,\cdots,\mathbf{n}_A) = :\rho_{i_1,j_1}(\mathbf{n}_1)\cdots\rho_{i_A,j_A}(\mathbf{n}_A):$, where $\rho_{i,j}(\mathbf{n})$ is the one-body density operator for nucleons with spin $i$ and isospin $j$ at lattice site $\mathbf{n}$ in coordinate space. A key feature of the pinhole method is that the exact locations of nucleons with spin $i$ and isospin $j$ can be sufficiently sampled by inserting the normal-ordered $A$-body density operator.
Then, we insert the operator $\rho_{i_1,j_1,\cdots,i_A,j_A}(\mathbf{n}_1,\cdots,\mathbf{n}_A)$ at middle time of the Euclidean time projection amplitudes,
\begin{equation}
\bra{{\Psi_f}} M^{L_t/2} \rho_{i_1,j_1,\cdots, i_A,j_A} (\mathbf{n}_1, \cdots, \mathbf{n}_A) M^{L_t/2} \ket{\Psi_i},
\label{eq:smeq2}
\end{equation}
where $\ket{\Psi_i}, \ket{\Psi_f}$ are the initial and final trial wave functions, propagated via the transfer matrix $M =:\exp(-H_{S} a_{t}):$. Here, $H_S$ denotes the simple Hamiltonian used in the propagation,
while $H_{\chi'}$ denotes the unitarily transformed full Hamiltonian obtained by wave-function matching, as discussed in Ref.~\cite{Elhatisari:2022zrb_sm}. 
The transfer matrix is connected with the evolution operator $\exp(-H_{S} \tau)$ via Trotterization, which approximates the time evolution by dividing it into discrete steps for computational feasibility. Here, the difference $H_{\chi}'-H_{S}$ is calculated by perturbation theory, see 
Ref.~\cite{Lu:2018bat_sm} for more details. In the large $\tau$ limit, we can obtain the required physical states. 
Due to the completeness relation of the $A$-body density operator $\rho_{i_1,j_1,\cdots,i_A,j_A}(\mathbf{n}_1,\cdots,\mathbf{n}_A)$, the sum of this expectation value over $\mathbf{n}_1,\cdots,\mathbf{n}_A$ and $i_1,j_1,\cdots,i_A,j_A$ yields $A!$ times the amplitude $\bra{\Psi_f} M^{L_t}\ket{\Psi_i}$, which can be calculated using Auxiliary Fields Monte Carlo simulations.
In the auxiliary field formalism, the transfer matrix $M$ at each time slice depends on the auxiliary fields $s_{n_{t}}$, and Eq. \eqref{eq:smeq2} can be rewritten in terms of the auxiliary fields. To evaluate it, we perform importance sampling according to the amplitude,
\begin{equation}
A(s;\mathbf{n}_1, \cdots, \mathbf{n}_A; i_1,j_1,\cdots, i_A,j_A) = |\bra{{\Psi_f}} M(s_{Lt})\cdots M(s_{Lt/2+1}) \rho_{i_1,j_1,\cdots, i_A,j_A} (\mathbf{n}_1, \cdots, \mathbf{n}_A)  M(s_{Lt/2})\cdots M(s_{1}) \ket{\Psi_i}|,
\end{equation}
which reflects the $A$-body density information with the auxiliary fields $s_{1},\cdots,s_{Lt}$, pinhole locations $\mathbf{n}_1,\cdots,\mathbf{n}_A$ and spin-isospins indices $i_1,j_1,\cdots,i_A,j_A$. To generate different amplitudes, the auxiliary fields $s_{1},\cdots,s_{Lt}$ are updated by the shuttle algorithm \cite{Lu:2018bat_sm}, and then the pinhole locations $\mathbf{n}_1,\cdots,\mathbf{n}_A$ and spin-isospins indices $i_1,j_1,\cdots,i_A,j_A$ are sampled by the  Metropolis algorithm. For any operator that does not mix pinhole configurations, we evaluate its expectation value as
\begin{equation}
\langle\hat{O}\rangle =\frac{\sum{A(s;\mathbf{n}_1, \cdots, \mathbf{n}_A; i_1,j_1,\cdots, i_A,j_A)} \mathrm{exp}(i\theta[s;\mathbf{n}_1, \cdots, \mathbf{n}_A; i_1,j_1,\cdots, i_A,j_A])O{(\mathbf{n}_1, \cdots, \mathbf{n}_A; i_1,j_1,\cdots, i_A,j_A)}}{\sum{ A(s;\mathbf{n}_1, \cdots, \mathbf{n}_A; i_1,j_1,\cdots, i_A,j_A)}\mathrm{exp}(i\theta[s;\mathbf{n}_1, \cdots, \mathbf{n}_A; i_1,j_1,\cdots, i_A,j_A])},
\end{equation}
where the summation symbol indicates the sum of different auxiliary fields $s$, pinhole locations $\mathbf{n}_1,\cdots,\mathbf{n}_A$ and spin-isospins indices $i_1,j_1,\cdots,i_A,j_A$, and $\mathrm{exp}(i\theta[s;\mathbf{n}_1, \cdots, \mathbf{n}_A; i_1,j_1,\cdots, i_A,j_A])$ is the complex phase of each pinhole amplitude.

\subsection{Intrinsic densities}
Recent experimental observations have provided growing evidence for intrinsic clustering in nuclei~\cite{Tanaka:2021oll_sm}. On the theoretical side, many-body approaches based on core or reference-state frameworks have achieved remarkable success in describing nuclear structure~\cite{Caurier:2004gf_sm,Hjorth-Jensen:1995zrg_sm,Coraggio:2008in_sm,Hagen:2013nca_sm,Hergert:2015awm_sm}. For the light nuclei studied in this work, it is therefore of particular interest to investigate the possible presence of preformed clusters and their role in shaping the intrinsic configurations of these systems. 
For each sampled pinhole configuration, we construct all nucleon triplets or quadruplets that satisfy the spin–isospin coupling required for the cluster. Since the clusters we aim to identify are expected to be relatively compact and approximately Gaussian in coordinate space~\cite{Tohsaki:2001an_sm,Funaki:2015uya_sm,Zhou:2019cjz_sm}, we assign to each candidate cluster a Gaussian-type probability weight based on the distances of its nucleons from the cluster center of mass (c.m.)
\begin{equation}
P \propto \exp\biggl[-\sum_i \bigl|\mathbf r_i - \mathbf r_{\text{c.m.}}\bigr|^2\biggr],
\end{equation}
where $\mathbf r_i$ denotes the position of the $i$th nucleon in the cluster and $\mathbf r_{\text{c.m.}}$ is the c.m. coordinate of the cluster.
In the present work, the candidate with the largest probability weight $P$ is selected as the intrinsic cluster, while more sophisticated selection schemes can be explored in future work.

\subsection{Schematic illustration of four-neutron correlations in $^{7}$H}

\begin{figure}[!h]
\centering
  \includegraphics[width=0.8\textwidth]{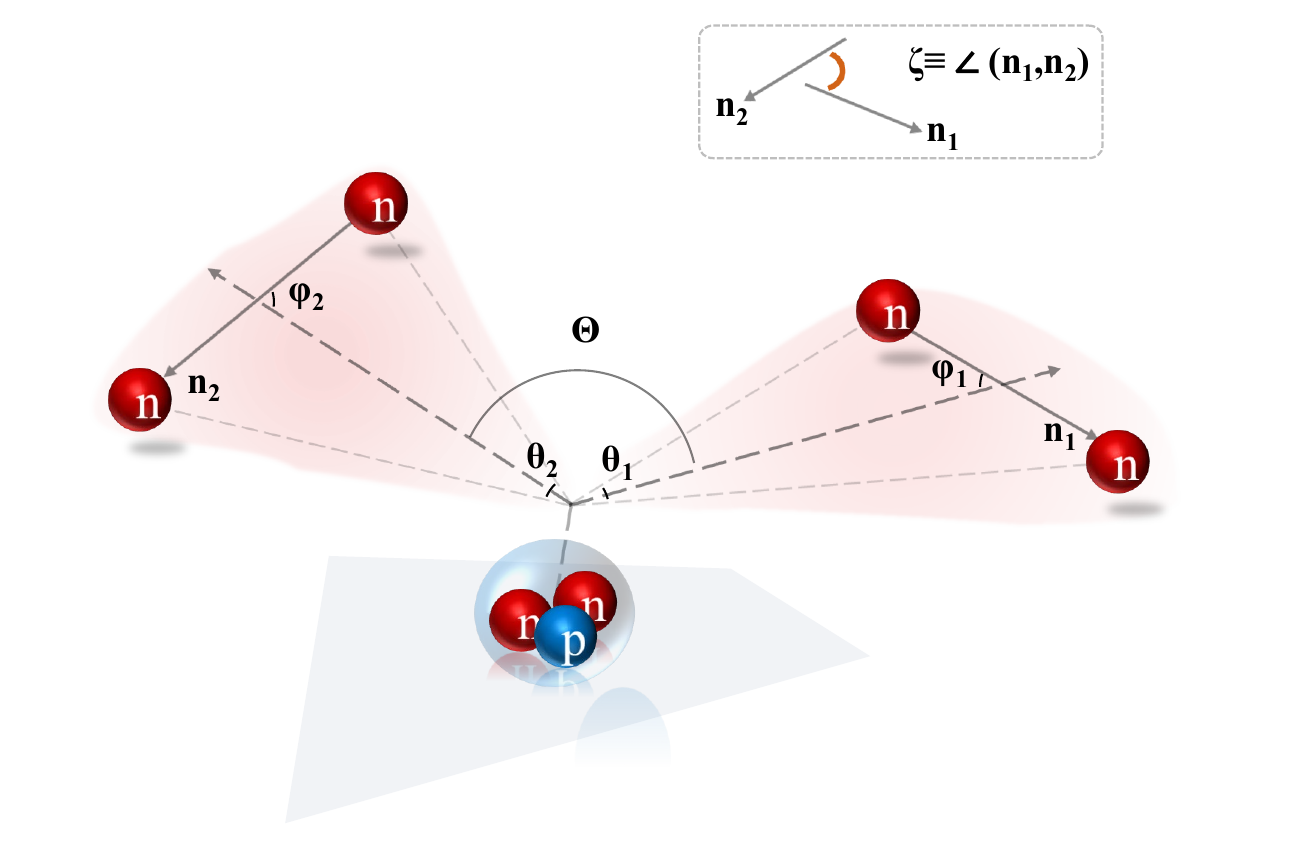}
  
  \caption{Schematic illustration of the angular coordinates for the reduced four-neutron correlations in $^{7}$H.}
  \label{sm0}
\end{figure}

Figure~\ref{sm0} schematically illustrates the angular coordinates used to characterize the four-body correlations in light nuclei, using $^{7}$H as an example.
The four neutrons are grouped into two neutron-neutron pairs, and for each pair we adopt the same two-body angular variables $(\theta,\varphi)$ as in the two-body correlation function. We define the pair vectors $\mathbf{n}_{1}$ and $\mathbf{n}_{2}$ as the relative displacement vectors connecting the two neutrons within each pair.
The intra-pair opening angles $\theta_{1}$ and $\theta_{2}$ quantify the compactness of the pairs, while $\varphi_{1}$ and $\varphi_{2}$ encode the degree of radial asymmetry between the two neutrons within a pair (as in the two-body analysis).
The angle $\Theta$ characterizes the relative opening between the pair c.m. vectors, and the torsion angle $\zeta \equiv \angle(\mathbf n_1,\mathbf n_2)
= \arccos\!\left(\frac{\left|\mathbf n_1\cdot \mathbf n_2\right|}{|\mathbf n_1|\,|\mathbf n_2|}\right)$ measures the non-coplanarity (twist) between the two neutron-neutron pair vectors.
This set of angles provides a convenient parameterization of the four-neutron geometry and underlies the angular distributions discussed in the main text.

\subsection{Extrapolation in Euclidean time}
Since it is computationally infeasible to evolve the Hamiltonian to infinite Euclidean time, an extrapolation to the limit $\tau\rightarrow\infty$ is required.
~Based on spectral decomposition of energy at Euclidean time $\tau$~\cite{Lahde:2014sla_sm,He:2019ipt_sm}, the expectation value of the Hamiltonian can be written as
\begin{equation}
E(\tau) = \frac{E_0 + \sum_{i}E_{i}c_{i}^2e^{-\Delta E_i \tau}}{1+ \sum_{i} c_{i}^2e^{-\Delta E_i \tau}},
\label{eq:eet1}
\end{equation}
where $i$ denotes the $i$-th eigenstate, $\Delta E_i$ is its energy gap to the g.s., and $c_{i}$ represents its overlap with the initial state relative to that of the g.s.
Following the procedure in Ref.~\cite{Shen:2023_sm}, we first fit the energies calculated with the simple Hamiltonian using the lowest order of Eq. \eqref{eq:eet1} to extract $\Delta E_i$ and $c_{i}$. These parameters are then used in a second-stage fit for general operators, such as higher-order energy correlations and observables that commute with the Hamiltonian, based on the functional form 
\begin{equation}
O(t) = \frac{O_0 + O_{1,1}e^{-\Delta E_1 \tau/2} + O_{1,2}e^{-\Delta E_1 \tau}}{1+  c_{1}^2e^{-\Delta E_1 \tau}}.
\end{equation}

\subsection{Finite-volume effects}
To assess finite-volume effects, we computed the ground-state energy of $^{7}$H on three lattice sizes $L = 11 (14.52~\text{fm})$, $12 (15.84~\text{fm})$, and $14 (18.48~\text{fm})$, using two fixed Euclidean times $\tau$ chosen to balance physical reliability and statistical uncertainty (see Table~\ref{tab:fve}).
The ground-state energies show a mild dependence on the lattice size. Specifically, the $L = 12$ results differ slightly from those at $L = 14$ by less than 0.05~MeV for both $\tau = 0.35$~MeV$^{-1}$ and $\tau = 0.45$~MeV$^{-1}$, and differ somewhat more from $L = 11$ by about 0.08~MeV, all within statistical uncertainties.
These findings indicate that finite-volume effects at $L = 12$ are minor, and no volume extrapolation is performed in this work. 
A more systematic treatment of finite-volume effects will be required in future calculations with higher statistical precision.

\begin{table}[htb]
\centering
    \setlength{\tabcolsep}{25pt}
\begin{tabular}{lccc}
\hline
$L$[fm] & $E$($\tau = 0.35~$MeV$^{-1}$)[MeV]& $E$($\tau = 0.45 ~$MeV$^{-1}$)[MeV] \\
\hline
14.52 & $-4.154 (106)$ & $-4.615 (208)$  \\\hline
15.84  & $-4.226 (113)$ & $-4.650 (194)$ \\\hline
18.48 & $-4.234 (129)$ & $-4.696 (284)$   \\\hline

\end{tabular}
\caption{Ground-state energy of the $^{7}$H calculated at different box sizes with the chiral force $H_{\chi}$. The errors in parentheses indicate the statistical uncertainties.}
\label{tab:fve}
\end{table}

\subsection{Experimental information on $^{6}$H and $^{7}$H}

Table~\ref{tab:H67_expt} summarizes experimental determinations of the energies and widths of $^{6}$H and $^{7}$H obtained with different reaction mechanisms. Energies $E$ and widths $\Gamma$ are given relative to the $^{3}$H+$3n$ threshold for $^{6}$H and the $^{3}$H+$4n$ threshold for $^{7}$H. 
For $^{6}$H and $^{7}$H, we list only those measurements that report candidate ground-state energies and, where available, widths. 
Other studies have provided evidence for the existence of these systems without extracting specific energies~\cite{Korsheninnikov:2003bz_sm}, and some experiments do not report any clear signal attributable to $^{6}$H or $^{7}$H under their experimental conditions~\cite{Parker:1990plb_sm,Seth:1991zz_sm,Evseev:1981dce_sm,Aleksandrov:1982cf_sm}.
The values measured in different reactions illustrate that the experimentally reported ground-state energies and widths of $^{6}$H and $^{7}$H differ significantly across reaction mechanisms and analyses.

\begin{table*}[h]
  \centering
  \setlength{\tabcolsep}{15pt}
  \caption{Experimental information on $^{6}$H and $^{7}$H. Energies $E$ and widths $\Gamma$ are given relative to the $^{3}$H+$3n$ threshold for $^{6}$H and the $^{3}$H+$4n$ threshold for $^{7}$H. The table highlights the spread of reported values across different reaction mechanisms and analyses.
}
  \label{tab:H67_expt}
  \begin{tabular}{clcc}
    \hline\hline
    Nucleus & $E$ (MeV) & $\Gamma$ (MeV)  & Exp. \\
    \hline
    $^{6}$H &
    $2.7 \pm 0.4$ & $1.8 \pm 0.5$ &
    $^{7}$Li($^{7}$Li,$^{8}$B)$^{6}$H~\cite{Aleksandrov:1984H6_sm} \\[0.2em]

    $^{6}$H &
    $2.6 \pm 0.5$ & $1.5 \pm 0.3$ &
    $^{9}$Be($^{11}$B,$^{14}$O)$^{6}$H~\cite{Belozyorov:1986H456_sm} \\[0.2em]
    
    $^{6}$H &
    $6.6 \pm 0.7$ & $5.5 \pm 2.0$ &
    $^{9}$Be($\pi^-$,$pd$)$^{6}$H~\cite{Gurov:2003pv_sm} \\[0.2em]

    $^{6}$H &
    $7.3 \pm 1.0$ &  $5.8 \pm 2.0$  &
    $^{11}$B($\pi^-$,$p$$^{4}$He)$^{6}$H~\cite{Gurov:2003pv_sm} \\[0.2em]

    $^{6}$H &
    $2.91^{+0.85}_{-0.95}$ & $1.52^{+1.77}_{-0.35}$ &
    $^{12}$C($^{8}$He,$^{14}$N)$^{6}$H~\cite{Caamano:2008zz_sm} \\[0.2em]

    $^{6}$H &
    $>4.5$ & -- &
    $^{2}$H($^{8}$He,$^{4}$He)$^{6}$H~\cite{Nikolskii:2021kqe_sm} \\[0.2em]

    $^{6}$H &
    $2.3 \pm 0.5\ \pm 0.4$ &
    $1.9 \pm 1.0 \pm 0.4$ &
    $^{7}$Li($e,e'p\pi^+$)$^{6}$H~\cite{A1:2025mjf_sm} \\[0.4em]
    
    \hline
    
    $^{7}$H &
    $\sim 1.7$ & $\leq$ 2.0 &
    p($^{8}$He,$^{2}$He)$^{7}$H~\cite{Korsheninnikov:2000He7unique_sm} \\[0.2em]

    $^{7}$H &
    $>0.05$--$0.10$ & -- &
    $^{2}$H($^{8}$He,$^{7}$H)$^{3}$He~\cite{Golovkov:2004lsg_sm} \\[0.2em]

    $^{7}$H &
    $0.57^{+0.42}_{-0.21}$ & $0.09^{+0.94}_{-0.02}$ &
    $^{12}$C($^{8}$He,$^{13}$N)$^{7}$H~\cite{Caamano:2007zz_sm} \\[0.2em]

    $^{7}$H &
    $\sim 2.0$ & -- &
    $^{2}$H($^{8}$He,$^{3}$He)$^{7}$H~\cite{Nikolskii:2010zz_sm} \\[0.2em]

    $^{7}$H &
    $1.8(5)$ & -- &
    $^{2}$H($^{8}$He,$^{3}$He)$^{7}$H~\cite{Bezbakh:2019dvh_sm} \\[0.2em]
    
    $^{7}$H &
    $0.73^{+0.58}_{-0.47}$ & $0.18^{+0.47}_{-0.16}$ &
    $^{12}$C($^{8}$He,$^{13}$N)$^{7}$H~\cite{Caamano:2020_sm} \\[0.2em]
    
    $^{7}$H &
    $2.2(5)$ & -- &
    $^{2}$H($^{10}$Be,$^{3}$He)$^{7}$H~\cite{Muzalevskii:2020svp_sm} \\[0.2em]
    
    \hline\hline
  \end{tabular}
\end{table*}

\begin{figure}[!h]
\centering
  \includegraphics[width=0.8\textwidth]{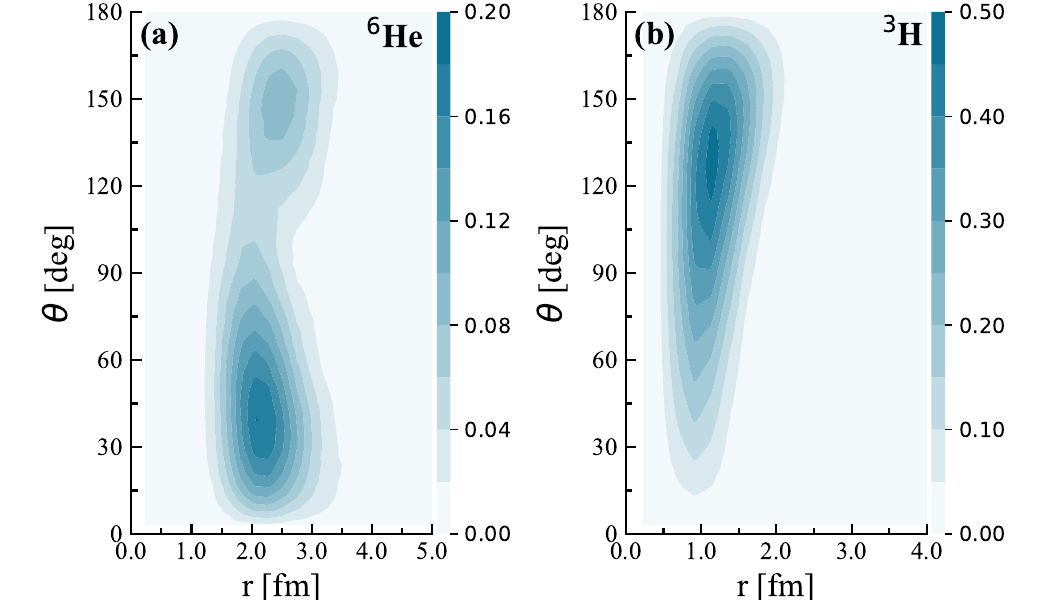}
  
  \caption{Two-neutron correlation densities of $^{6}$He and $^{3}$H. (a) Two-neutron correlation density of the valence neutrons in $^{6}$He, used as a benchmark for our correlation analysis. (b) Two-neutron correlation density in $^{3}$H, shown as a reference for comparison with the short-distance component in $^{7}$H.}
  \label{sm1}
\end{figure}

\subsection{Two-neutron correlations in $^{6}$He and $^{3}$H}

Figure~\ref{sm1}(a) shows the two-neutron correlation density of the valence neutrons in $^{6}$He. 
Since $^{6}$He has been extensively studied both experimentally and theoretically~\cite{Hagino:2005we_sm,Papadimitriou:2011jx_sm,Assie:2010zz_sm}, it provides a benchmark for our correlation analysis.
Figure~\ref{sm1}(b) shows the two-neutron correlation density of $^{3}$H for comparison with $^{7}$H.
Note that the $^{7}$H correlations are normalized by the total number of neutron pairs, $\binom{6}{2}=15$. 
Therefore, the per-pair short-distance component shown for $^{7}$H is reduced by a factor of 15 relative to the $^{3}$H reference purely due to this normalization convention.
The two-neutron correlation density characterizes the spatial extent and angular structure of the neutron pair in $^{3}$H and closely resembles the short-distance component of the multi-neutron correlation patterns obtained in $^{7}$H.

\begin{figure}[h]
\centering
  \includegraphics[width=0.8\textwidth]{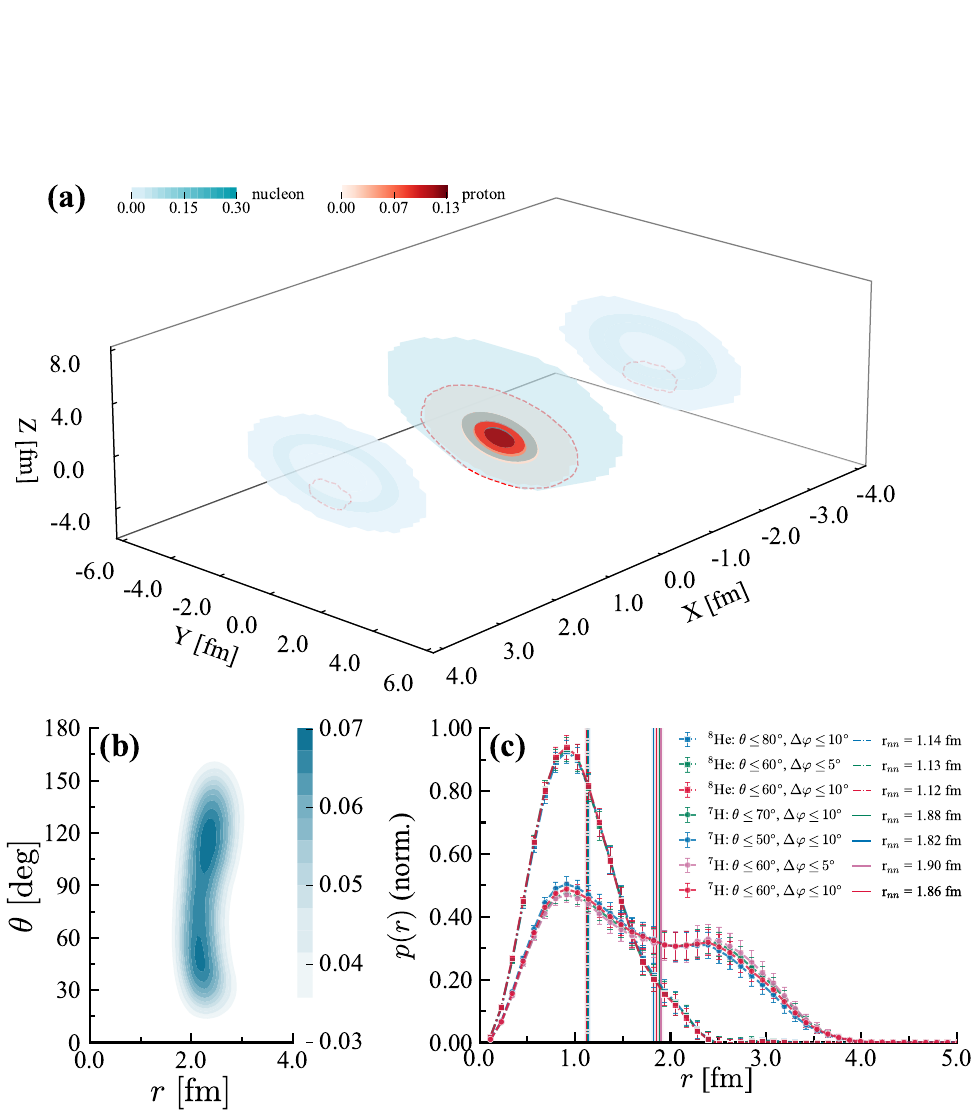}
  
  \caption{(a) Intrinsic nucleon (blue) and proton (red) densities of the $^8$He ground state, visualized as slices at $x=0$ and $x=\pm 3$~fm. The red dashed line marks the maximal spatial extent of the identified $\alpha$ cluster. (b)  Two-neutron correlation density among four valence neutrons in the $^8$He ground state, where $r$ denotes the distance of each neutron from the $^{8}$He c.m. and $\theta$ is the opening angle between the two neutron positions. (c) Normalized dineutron size ( $r\equiv \tfrac12|\mathbf r_1-\mathbf r_2|$) probability density $p(r)$ (in~fm$^{-1}$) for dineutron-like configurations in $^8$He and $^7$H with different angular selection windows in $(\theta, \Delta\varphi)$, where $\Delta\varphi \equiv |\varphi-90^\circ|$ quantifies the deviation from radial symmetry.
}
  \label{sm2}
\end{figure}

\subsection{Intrinsic densities, two-neutron correlations, and dineutron sizes in $^{8}$He and $^{7}$H}

Figure~\ref{sm2} summarizes the intrinsic nucleon densities, two-neutron correlations, and resulting dineutron-size distributions for $^{8}$He and $^{7}$H, analyzed within the same intrinsic-frame definitions as in the main text.
Panel~(a) shows the intrinsic total-nucleon and proton densities, visualized as slices at $x = 0$ and $x = \pm 3$~fm.
The intrinsic densities indicate a compact $\alpha$-like cluster, surrounded by four valence neutrons, with a more compact valence-neutron distribution than in $^{7}$H.
Panel~(b) displays the two-neutron correlation density for the four valence neutrons in $^{8}$He, where $r$ denotes the distance of each neutron from the $^{8}$He c.m. and $\theta$ is the opening angle between the two neutron positions.
When the two-neutron correlations are averaged over all valence-neutron pairs in $^{8}$He, the correlation density exhibits a single broad maximum around $\theta\simeq 110^\circ$ in our calculation.
In the idealized limit of a pure $(0p_{3/2})^4$ shell-model configuration, this averaging largely washes out orientation dependence, yielding an approximately isotropic neutron-neutron angular density. Consequently, the resulting distribution is dominated by the phase-space factor $\sin\theta$.
The contrast with $^{7}$H is instructive: while both systems deviate from the idealized $(0p_{3/2})^4$ limit, the departure is substantially stronger in $^{7}$H, allowing nontrivial angular structures to survive the pair averaging, whereas $^{8}$He remains comparatively closer to a dominant $(0p_{3/2})^4$ configuration.
In this analysis, we focus on valence-neutron pairs with opposite spin projections. This selection suppresses the largely uncorrelated background (including cross pairs spanning different correlated clusters) and enhances spatially clustered configurations.
As shown in panel~(b), this conditional correlation exhibits a characteristic ridge at small opening angles $\theta$, signaling compact dineutron-like configurations, together with a pronounced enhancement toward large $\theta$, consistent with the bimodal pattern discussed for $^{7}$H.
Panel~(c) compares the normalized dineutron size distributions for dineutron-like configurations in $^{8}$He and $^{7}$H selected by combined cuts on the internal opening angle $\theta$ and the angle $\varphi$ (see legend for the specific cut choices). 
Here, we define the dineutron size as the half-separation $r\equiv \tfrac12|\mathbf r_1-\mathbf r_2|$, whereas the dineutron size often quoted in the literature refers to the full relative distance $|\mathbf r_1-\mathbf r_2|=2r$.
From these distributions we extract the root-mean-square dineutron radius, $ r_{nn} = \sqrt{\int dr\, r^2\,\rho(r)}$, which is indicated by the vertical lines and quoted in the legend. 
The extracted $r_{nn}$ shows only a weak dependence on the $(\theta,\Delta\varphi)$ cuts and remains consistent within uncertainties, indicating that the inferred dineutron size is stable against the selection procedure. This also suggests that our choice of angular windows lies in a reasonable range.
Moreover, the extracted dineutron size in $^{8}$He corresponds to a typical dineutron size of $2.24(4)$~fm, consistent with the commonly considered range to be $2–3$~fm~\cite{Zhukov:1993aw_sm,Pillet:2007hb_sm} at the nuclear surface, whereas in 
$^{7}$H it exhibits a larger typical scale of $3.72(8)$~fm.

\subsection{Opening angle $\theta$ and $\varphi$ dependence in four-neutron correlation of $^{7}$H}

\begin{figure}[h]
\centering
  \includegraphics[width=\textwidth]{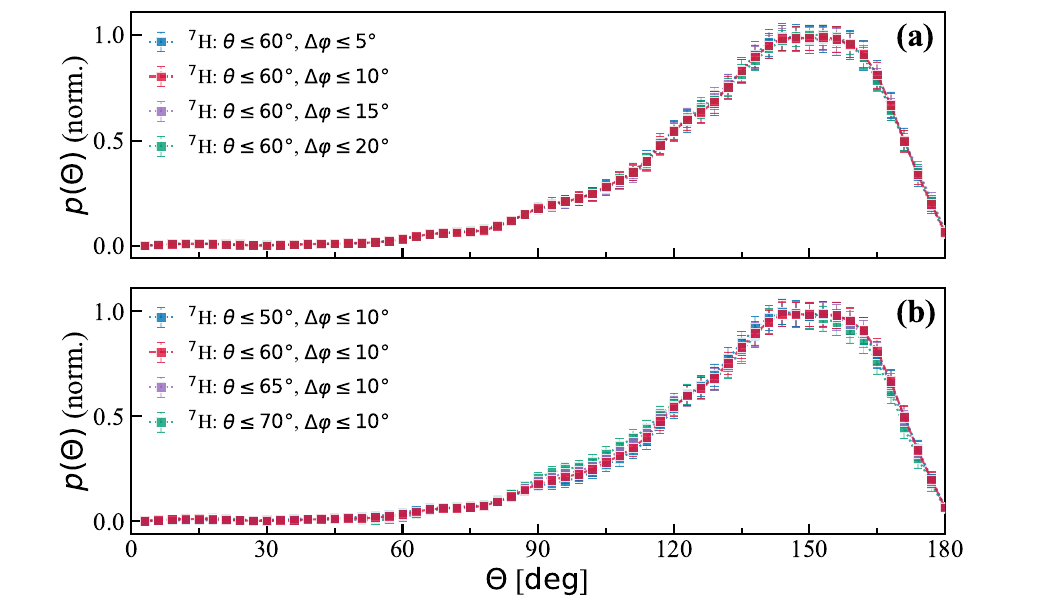}
  
  \caption{Robustness of the four-neutron inter-pair opening-angle distribution to intra-pair selections in $^{7}$H.
  (a) Normalized probability densities of the $\Theta$ under a fixed intra-pair selection $\theta \le 60^\circ$, comparing four $\Delta\varphi$ selections: $\Delta \varphi \le 5^\circ, 10^\circ, 15^\circ$, and $20^\circ$.
(b) Same as (a) for fixed $\Delta\varphi \le 10^\circ$ while varying the neutron-neutron opening-angle cut $\theta \le 50^\circ, 60^\circ, 65^\circ$, and $70^\circ$.
Angles are shown in degrees, while densities are reported per radian.
}
  \label{sm3}
\end{figure}

We assess the robustness of the inter-pair opening-angle distribution $p(\Theta)$ to variations of the intra-pair angle selections used to identify dineutron-like neutron pairs  in Figs.~\ref{sm3}.
Here $\Theta$ is the inter-pair opening angle between the two pair c.m. position vectors (measured from the system c.m.). The intra-pair geometry is characterized by the neutron--neutron opening angle $\theta$ and by $\Delta\varphi\equiv|\varphi-90^\circ|$, which quantifies the deviation from radial symmetry (where $\varphi$ is defined as the angle between the relative position vector of the two neutrons and the bisector of their opening angle).
As shown in Fig.~\ref{sm3}(a), changing $\Delta\varphi$ from $5^\circ$ to $20^\circ$ at fixed $\theta \le 60^\circ$ leaves the normalized $\Theta$ distribution essentially unchanged.
When we vary the $\Delta\varphi$ selection window, it primarily changes the accepted sample size (i.e., the number of accepted configurations) but leaves the normalized shape of the $\Theta$ distribution essentially unchanged, indicating no discernible change in the underlying four-neutron correlation pattern within uncertainties. In this work, we select $\Delta\varphi \leq 10^\circ$ to balance the radial symmetry and statistics.
In Fig.~\ref{sm3}(b), we vary the intra-pair neutron-neutron opening-angle cut $\theta$ while keeping $\Delta\varphi \le 10^\circ$ fixed.
Because $\theta$ directly quantifies the internal geometry of a neutron-neutron pair, this cut defines which intra-pair correlation patterns are included in the sampled configurations.
Guided by the two-body correlation analysis discussed above, we find that the dominant strength associated with compact, dineutron-like correlations in $^{7}$H is concentrated primarily within $\theta \le 60^\circ$, which motivates our baseline choice.
Accordingly, for moderate variations of the $\theta$ cut ($50^\circ$--$70^\circ$), the normalized $\Theta$ distribution exhibits only minor differences and remains consistent within uncertainties.
This indicates that our baseline selection captures the relevant neutron-neutron correlation physics without either overly truncating compact configurations or introducing spurious distortions.

\begin{figure}[h]
\centering
  \includegraphics[width=0.8\textwidth]{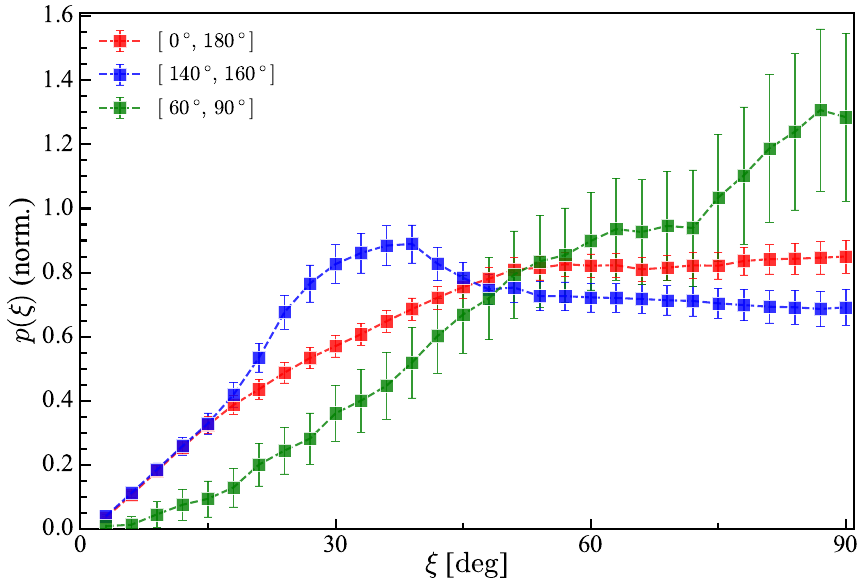}
  
  \caption{Normalized probability densities $p(\xi)$ of the dihedral angle $\xi$ between the two planes spanned by each neutron pair and the c.m. of $^{7}$H. Angles are in degrees, and densities are reported per radian. The selection windows $\theta \leq 60^\circ$ and $\Delta\varphi \leq 10^\circ$ are identical to those used in the analyses of the torsion angle and inter-pairing opening angle. The three curves correspond to different cuts on the inter-pair opening angle:  $\Theta\in[0^\circ,180^\circ]$,  $\Theta\in[140^\circ,160^\circ]$, and $\Theta\in[60^\circ,90^\circ]$.
}
  \label{sm4}
\end{figure}
\subsection{Distribution of the dihedral angle in $^{7}$H}

To quantify the non-coplanarity of four-neutron configurations in $^{7}$H, we introduce the dihedral angle $\xi$ between the two planes spanned by each neutron pair and the c.m. of $^{7}$H. 
Using the neutron coordinates relative to the $^{7}$H c.m., we construct the normal vectors of two pair-c.m. planes for each pinhole configuration. 
By construction, $\xi \approx 0^\circ$ corresponds to approximately coplanar arrangements of the two planes. 
As $\xi$ increases, the planes open up and approach an orthogonal configuration near $\xi \approx 90^\circ$. 
Because a plane normal is defined only up to a sign (depending on the orientation convention of the in-plane vectors used to construct it), the dihedral angles $\xi$ and $180^\circ-\xi$ are physically equivalent. 
We therefore fold the distribution and restrict the dihedral angle to $0^\circ \le \xi \le 90^\circ$ without loss of information.
Figure~\ref{sm4} shows the normalized probability density $p(\xi)$ for different selections of the inter-pair opening angle $\Theta$. 
For the inclusive selection $\Theta\in[0^\circ,180^\circ]$, $p(\xi)$ is relatively broad, indicating that the sampled configurations span a wide range of torsional geometries, while the weight near the coplanar limit ($\xi\approx 0^\circ$) is suppressed. 
For two selected intervals, the distributions for $\Theta\in[140^\circ,160^\circ]$ exhibit a pronounced peak at intermediate dihedral angles, $\xi\approx 30^\circ$-$40^\circ$, whereas those for $\Theta\in[60^\circ,90^\circ]$ shift strength toward larger dihedral angles and peak near $\xi\approx 90^\circ$.

\subsection{Internal angular and radial distributions from the four-neutron correlations in $^{7}$H}

\begin{figure}[h]
\centering
  \includegraphics[width=\textwidth]{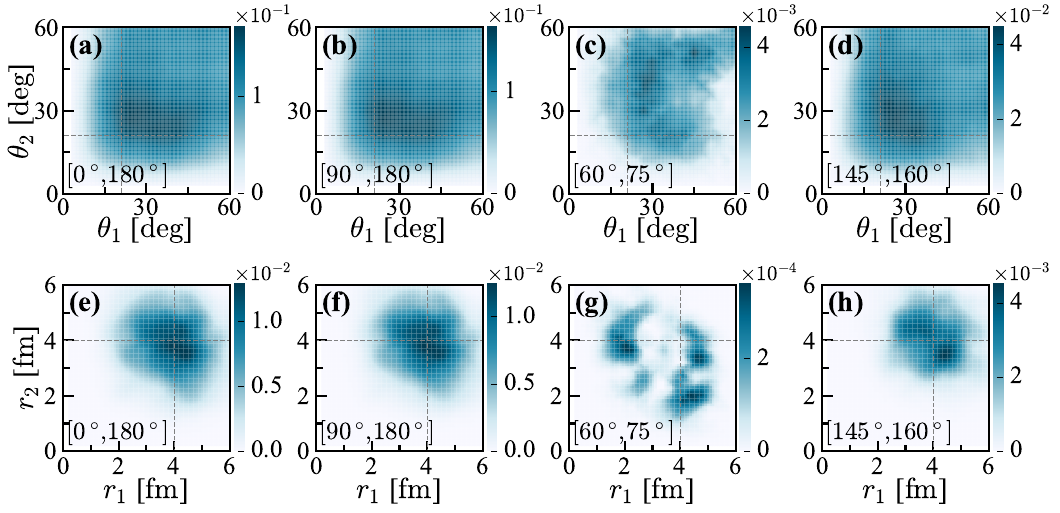}
  
\caption{
Internal angular and radial distributions from the four-neutron correlation analysis in $^{7}$H.
(a--d) Joint probability density $p(\theta_1,\theta_2)$ of the intra-pair opening angles of two opposite-spin $nn$ pairs defined in each sampled four-neutron configuration.
(e--h) Corresponding joint probability density $p(r_1,r_2)$ of the pair radial coordinates, where $r_i$ denotes the distance from the $i$th pair c.m. to the $^{7}$H c.m. With the $\Delta\varphi$ selection favoring the two neutrons in each pair to be approximately equidistant from the $^{7}$H c.m., the pair c.m.\ radius $r_i$ provides a convenient measure of the radial location.
Columns correspond to selections on the inter-pair opening angle $\Theta$ (indicated in each panel):
$\Theta\in[0^\circ,180^\circ]$ (a,e),
$\Theta\in[90^\circ,180^\circ]$ (b,f),
and two characteristic four-neutron configuration windows, $\Theta\in[60^\circ,75^\circ]$ (c,g) and $\Theta\in[145^\circ,160^\circ]$ (d,h), where the marginal distribution $p(\Theta)$ is enhanced.
Dashed lines mark the typical dineutron scales (intra-pair opening angles and pair radial locations) extracted independently from the two-neutron correlation analysis, indicating that the dominant internal scales inferred from two-neutron correlations are consistent with those extracted from the explicit four-neutron correlations.
}
  \label{sm5}
\end{figure}

Figure~\ref{sm5} provides additional information on the internal angular and radial distributions in the four-neutron correlation analysis and serves as a consistency check against the two-neutron correlations in $^{7}$H.
Panels (a,e) compile the joint probability densities $p(\theta_1,\theta_2)$ and $p(r_1,r_2)$ for the two neutron-neutron pairs identified in each sampled four-neutron configuration.
The bulk of the probability density is concentrated around the typical dineutron angular and radial scales extracted independently from the two-neutron correlations (dashed lines), indicating that the leading internal scales inferred from two-body observables are consistent with those obtained from the explicit four-body correlation.
This agreement supports the interpretation of the observed bimodal angular
pattern in $^{7}$H from the two-body correlation analysis: the two-body correlation function represents a statistical distribution over all neutron-neutron pairs and thus provides a probe sensitive to the underlying multi-neutron geometry.
Panels (b,f), obtained with the selection $\Theta\in[90^\circ,180^\circ]$, are similar to (a,e) and contain the majority of the sampled configurations, indicating that the $\Theta>90^\circ$ (extended) sector dominates the sampled four-body ensemble.
Panels (d,h) correspond to a characteristic $\Theta$ window around a maxima of the marginal distribution $p(\Theta)$ ($\Theta\in[145^\circ,160^\circ]$).
For the large-$\Theta$ window (d,h), the radial distribution $p(r_1,r_2)$ shows that $r_1$ and $r_2$ are comparable, and $p(\theta_1,\theta_2)$ shifts toward smaller intra-pair opening angles, consistent with a near back-to-back arrangement of the two pair c.m. vectors at comparable radii and a more compact intra-pair geometry.
In contrast, the small-$\Theta$ window (c,g) shows a more asymmetric joint radial distribution, featuring enhanced weight near $(r_1,r_2)\approx(2,4)$~fm, alongside a nearly symmetric contribution at $r_1\approx r_2\approx4$~fm. The intra-pair opening-angle distribution shifts to larger angles and becomes noticeably broader, with enhanced support around $60^\circ$, compared to the tighter concentration near $30^\circ$ in panel (d).
The associated torsion- and dihedral-angle distributions are also weighted toward large angles in our calculation, indicating a strongly noncoplanar and interlaced four-neutron geometry with pronounced four-body correlations.

%


\begin{thebibliography}{40}%
\makeatletter

\bibitem{Marques:2021mqf}
F.~M.~Marqu{\'e}s and J.~Carbonell,
Eur. Phys. J. A \textbf{57} (2021) no.3, 105
doi:10.1140/epja/s10050-021-00417-8
[arXiv:2102.10879 [nucl-ex]].

\bibitem{Marques:2001wh}
F.~M.~Marques, M.~Labiche, N.~A.~Orr, J.~C.~Angelique, L.~Axelsson, B.~Benoit, U.~C.~Bergmann, M.~J.~G.~Borge, W.~N.~Catford and S.~P.~G.~Chappell, \textit{et al.}
Phys. Rev. C \textbf{65} (2002), 044006
doi:10.1103/PhysRevC.65.044006
[arXiv:nucl-ex/0111001 [nucl-ex]].


\bibitem{Kisamori:2016jie}
K.~Kisamori, S.~Shimoura, H.~Miya, S.~Michimasa, S.~Ota, M.~Assie, H.~Baba, T.~Baba, D.~Beaumel and M.~Dozono, \textit{et al.}
Phys. Rev. Lett. \textbf{116} (2016) no.5, 052501
doi:10.1103/PhysRevLett.116.052501


\bibitem{Duer:2022ehf}
M.~Duer, T.~Aumann, R.~Gernh{\"a}user, V.~Panin, S.~Paschalis, D.~M.~Rossi, N.~L.~Achouri, D.~Ahn, H.~Baba and C.~A.~Bertulani, \textit{et al.}
Nature \textbf{606} (2022) no.7915, 678-682
doi:10.1038/s41586-022-04827-6

\bibitem{Huang:2021lch}
S.~W.~Huang, Z.~H.~Yang, F.~M.~Marqu{\'e}s, N.~L.~Achouri, D.~S.~Ahn, T.~Aumann, H.~Baba, D.~Beaumel, M.~B{\"o}hmer and K.~Boretzky, \textit{et al.}
Few Body Syst. \textbf{62} (2021) no.4, 102
doi:10.1007/s00601-021-01691-4

\bibitem{Oertel:2016bki}
M.~Oertel, M.~Hempel, T.~Kl{\"a}hn and S.~Typel,
Rev. Mod. Phys. \textbf{89} (2017) no.1, 015007
doi:10.1103/RevModPhys.89.015007
[arXiv:1610.03361 [astro-ph.HE]].

\bibitem{Ye:2024slx}
Y.~L.~Ye, X.~F.~Yang, H.~Sakurai and B.~S.~Hu,
Nature Rev. Phys. \textbf{7} (2025) no.1, 21-37
doi:10.1038/s42254-024-00782-5

\bibitem{Hen:2016kwk}
O.~Hen, G.~A.~Miller, E.~Piasetzky and L.~B.~Weinstein,
Rev. Mod. Phys. \textbf{89} (2017) no.4, 045002
doi:10.1103/RevModPhys.89.045002
[arXiv:1611.09748 [nucl-ex]].

\bibitem{Arrington:2022sov}
J.~Arrington, N.~Fomin and A.~Schmidt,
Ann. Rev. Nucl. Part. Sci. \textbf{72} (2022), 307-337
doi:10.1146/annurev-nucl-102020-022253
[arXiv:2203.02608 [nucl-ex]].


\bibitem{Tanihata:1985psr}
I.~Tanihata, H.~Hamagaki, O.~Hashimoto, Y.~Shida, N.~Yoshikawa, K.~Sugimoto, O.~Yamakawa, T.~Kobayashi and N.~Takahashi,
Phys. Rev. Lett. \textbf{55} (1985), 2676-2679
doi:10.1103/PhysRevLett.55.2676

\bibitem{Nakamura:2006zz}
T.~Nakamura, A.~M.~Vinodkumar, T.~Sugimoto, N.~Aoi, H.~Baba, D.~Bazin, N.~Fukuda, T.~Gomi, H.~Hasegawa and N.~Imai, \textit{et al.}
Phys. Rev. Lett. \textbf{96} (2006), 252502
doi:10.1103/PhysRevLett.96.252502

\bibitem{Tanihata:2013jwa}
I.~Tanihata, H.~Savajols and R.~Kanungo,
Prog. Part. Nucl. Phys. \textbf{68} (2013), 215-313
doi:10.1016/j.ppnp.2012.07.001

\bibitem{Kikuchi:2016ofi}
Y.~Kikuchi, K.~Ogata, Y.~Kubota, M.~Sasano and T.~Uesaka,
PTEP \textbf{2016} (2016) no.10, 103D03
doi:10.1093/ptep/ptw148
[arXiv:1603.03858 [nucl-th]].

\bibitem{R3B:2018lfr}
A.~Revel \textit{et al.} [R3B],
Phys. Rev. Lett. \textbf{120} (2018) no.15, 152504
doi:10.1103/PhysRevLett.120.152504
[arXiv:1803.04777 [nucl-ex]].

\bibitem{Kubota:2020pxo}
Y.~Kubota, A.~Corsi, G.~Authelet, H.~Baba, C.~Caesar, D.~Calvet, A.~Delbart, M.~Dozono, J.~Feng and F.~Flavigny, \textit{et al.}
Phys. Rev. Lett. \textbf{125} (2020), 252501
doi:10.1103/PhysRevLett.125.252501
[arXiv:2010.04802 [nucl-ex]].


\bibitem{Corsi:2023dek}
A.~Corsi, Y.~Kubota, J.~Casal, M.~G{\'o}mez-Ramos, A.~M.~Moro, G.~Authelet, H.~Baba, C.~Caesar, D.~Calvet and A.~Delbart, \textit{et al.}
Phys. Lett. B \textbf{840} (2023), 137875
doi:10.1016/j.physletb.2023.137875
[arXiv:2307.06083 [nucl-ex]].

\bibitem{Monteagudo:2024obw}
B.~Monteagudo, F.~M.~Marqu{\'e}s, J.~Gibelin, N.~A.~Orr, A.~Corsi, Y.~Kubota, J.~Casal, J.~G{\'o}mez-Camacho, G.~Authelet and H.~Baba, \textit{et al.}
Phys. Rev. Lett. \textbf{132} (2024) no.8, 082501
doi:10.1103/PhysRevLett.132.082501
[arXiv:2401.16817 [nucl-ex]].

\bibitem{Hammen:2018iny}
M.~Hammen, W.~N{\"o}rtersh{\"a}user, D.~L.~Balabanski, M.~L.~Bissell, K.~Blaum, I.~Budin{\v{c}}evi{\'c}, B.~Cheal, K.~T.~Flanagan, N.~Fr{\"o}mmgen and G.~Georgiev, \textit{et al.}
Phys. Rev. Lett. \textbf{121} (2018) no.10, 102501
doi:10.1103/PhysRevLett.121.102501

\bibitem{Geldhof:2022}
S.~Geldhof, M.~Kortelainen, O.~Beliuskina, P.~Campbell, L.~C\'aceres, L.~Ca{\~n}ete, B.~Cheal, K.~Chrysalidis, C.~S.~Devlin, R.~P.~de~Groote, A.~de~Roubin, T.~Eronen, Z.~Ge, W.~Gins, A.~Koszorus, S.~Kujanp\"a\"a, D.~Nesterenko, A.~Ortiz-Cortes, I.~Pohjalainen, I.~D.~Moore, A.~Raggio, M.~Reponen, J.~Romero and F.~Sommer,
Phys.\ Rev.\ Lett.\ \textbf{128} (2022) no.15, 152501
doi:10.1103/PhysRevLett.128.152501


\bibitem{Miller2019}
A.~J.~Miller, K.~Minamisono, A.~Klose \textit{et al.},
Proton superfluidity and charge radii in proton-rich calcium isotopes,
Nat.\ Phys.\ \textbf{15}, 432--436 (2019).

\bibitem{SAMURAI21-NeuLAND:2024kah}
J.~Kahlbow \textit{et al.} [SAMURAI21-NeuLAND],
Phys. Rev. Lett. \textbf{133} (2024) no.8, 082501
doi:10.1103/PhysRevLett.133.082501
[arXiv:2407.19303 [nucl-ex]].

\bibitem{Blank:2007zz}
B.~Blank and M.~Ploszajczak,
Rept. Prog. Phys. \textbf{71} (2008), 046301
doi:10.1088/0034-4885/71/4/046301
[arXiv:0709.3797 [nucl-ex]].

\bibitem{Pfutzner:2023tvr}
M.~Pf{\"u}tzner, I.~Mukha and S.~M.~Wang,
Prog. Part. Nucl. Phys. \textbf{132} (2023), 104050
doi:10.1016/j.ppnp.2023.104050
[arXiv:2304.13391 [nucl-ex]].


\bibitem{Pfutzner:2011ju}
M.~Pfutzner, M.~Karny, L.~V.~Grigorenko and K.~Riisager,
Rev. Mod. Phys. \textbf{84} (2012), 567
doi:10.1103/RevModPhys.84.567
[arXiv:1111.0482 [nucl-ex]].


\bibitem{Spyrou:2012zz}
A.~Spyrou, Z.~Kohley, T.~Baumann, D.~Bazin, B.~A.~Brown, G.~Christian, P.~A.~DeYoung, J.~E.~Finck, N.~Frank and E.~Lunderberg, \textit{et al.}
Phys. Rev. Lett. \textbf{108} (2012), 102501
doi:10.1103/PhysRevLett.108.102501

\bibitem{Kohley:2013kxt}
Z.~Kohley, T.~Baumann, D.~Bazin, G.~Christian, P.~A.~DeYoung, J.~E.~Finck, N.~Frank, M.~Jones, E.~Lunderberg and B.~Luther, \textit{et al.}
Phys. Rev. Lett. \textbf{110} (2013) no.15, 152501
doi:10.1103/PhysRevLett.110.152501
[arXiv:1303.2617 [nucl-ex]].






\bibitem{Tanaka:2021oll}
J.~Tanaka, Z.~H.~Yang, S.~Typel, S.~Adachi, S.~W.~Bai, P.~van Beek, D.~Beaumel, Y.~Fujikawa, J.~X.~Han and S.~Heil, \textit{et al.}
Science \textbf{371} (2021) no.6526, 260-264
doi:10.1126/science.abe4688


\bibitem{Freer:2017gip}
M.~Freer, H.~Horiuchi, Y.~Kanada-En'yo, D.~Lee and U.-G.~Mei{\ss}ner,
Rev. Mod. Phys. \textbf{90} (2018) no.3, 035004
doi:10.1103/RevModPhys.90.035004
[arXiv:1705.06192 [nucl-th]].

\bibitem{Kondo:2023F}
Y.~Kondo, N.~L.~Achouri, H.~Al~Falou, L.~Atar, T.~Aumann, H.~Baba, K.~Boretzky, C.~Caesar, D.~Calvet, H.~Chae \textit{et al.},
Nature \textbf{620} (2023) no.7976, 965--970
doi:10.1038/s41586-023-06352-6

\bibitem{Bezbakh:2019dvh}
A.~A.~Bezbakh, V.~Chudoba, S.~A.~Krupko, S.~G.~Belogurov, D.~Biare, A.~S.~Fomichev, E.~M.~Gazeeva, A.~V.~Gorshkov, L.~V.~Grigorenko and G.~Kaminski, \textit{et al.}
Phys. Rev. Lett. \textbf{124} (2020) no.2, 022502
doi:10.1103/PhysRevLett.124.022502
[arXiv:1906.07818 [nucl-ex]].

\bibitem{Muzalevskii:2020svp}
I.~A.~Muzalevskii, A.~A.~Bezbakh, E.~Y.~Nikolskii, V.~Chudoba, S.~A.~Krupko, S.~G.~Belogurov, D.~Biare, A.~S.~Fomichev, E.~M.~Gazeeva and A.~V.~Gorshkov, \textit{et al.}
Phys. Rev. C \textbf{103} (2021) no.4, 044313
doi:10.1103/PhysRevC.103.044313
[arXiv:2010.09655 [nucl-ex]].


\bibitem{A1:2025mjf}
T.~Shao \textit{et al.} [A1],
Phys. Rev. Lett. \textbf{134} (2025) no.16, 162501
doi:10.1103/PhysRevLett.134.162501
[arXiv:2501.01232 [nucl-ex]].

\bibitem{Grigorenko:2003mc}
L.~V.~Grigorenko and M.~V.~Zhukov,
Phys. Rev. C \textbf{68} (2003), 054005
doi:10.1103/PhysRevC.68.054005


\bibitem{Hagino:2005we}
K.~Hagino and H.~Sagawa,
Phys. Rev. C \textbf{72} (2005), 044321
doi:10.1103/PhysRevC.72.044321
[arXiv:nucl-th/0508058 [nucl-th]].

\bibitem{Hagino:2006ib}
K.~Hagino, H.~Sagawa, J.~Carbonell and P.~Schuck,
Phys. Rev. Lett. \textbf{99} (2007), 022506
doi:10.1103/PhysRevLett.99.022506
[arXiv:nucl-th/0611064 [nucl-th]].


\bibitem{Papadimitriou:2011jx}
G.~Papadimitriou, A.~T.~Kruppa, N.~Michel, W.~Nazarewicz, M.~Ploszajczak and J.~Rotureau,
Phys. Rev. C \textbf{84} (2011), 051304
doi:10.1103/PhysRevC.84.051304
[arXiv:1109.0223 [nucl-th]].

\bibitem{Grigorenko:2013uqa}
L.~V.~Grigorenko, I.~G.~Mukha and M.~V.~Zhukov,
Phys. Rev. Lett. \textbf{111} (2013) no.4, 042501
doi:10.1103/PhysRevLett.111.042501
[arXiv:1304.4901 [nucl-th]].

\bibitem{Kruppa:2013ala}
A.~T.~Kruppa, G.~Papadimitriou, W.~Nazarewicz and N.~Michel,
Phys. Rev. C \textbf{89} (2014) no.1, 014330
doi:10.1103/PhysRevC.89.014330
[arXiv:1310.7519 [nucl-th]].

\bibitem{Romero-Redondo:2016qmc}
C.~Romero-Redondo, S.~Quaglioni, P.~Navratil and G.~Hupin,
Phys. Rev. Lett. \textbf{117} (2016) no.22, 222501
doi:10.1103/PhysRevLett.117.222501
[arXiv:1606.00066 [nucl-th]].

\bibitem{Wang:2017qij}
S.~M.~Wang, N.~Michel, W.~Nazarewicz and F.~R.~Xu,
Phys. Rev. C \textbf{96} (2017) no.4, 044307
doi:10.1103/PhysRevC.96.044307
[arXiv:1707.08954 [nucl-th]].

\bibitem{Ma:2020roi}
Y.~Z.~Ma, F.~R.~Xu, N.~Michel, S.~Zhang, J.~G.~Li, B.~S.~Hu, L.~Coraggio, N.~Itaco and A.~Gargano,
Phys. Lett. B \textbf{808} (2020), 135673
doi:10.1016/j.physletb.2020.135673
[arXiv:2008.01420 [nucl-th]].

\bibitem{Bertulani:2008gq}
C.~A.~Bertulani, M.~S.~Hussein and G.~Verde,
Phys. Lett. B \textbf{666} (2008), 86-90
doi:10.1016/j.physletb.2008.06.062
[arXiv:0801.0784 [nucl-th]].

\bibitem{Delion:2013qwa}
D.~S.~Delion, R.~J.~Liotta and R.~Wyss,
Phys. Rev. C \textbf{87} (2013) no.3, 034328
doi:10.1103/PhysRevC.87.034328

\bibitem{Oishi:2014dna}
T.~Oishi, K.~Hagino and H.~Sagawa,
Phys. Rev. C \textbf{90} (2014) no.3, 034303
doi:10.1103/PhysRevC.90.034303
[arXiv:1404.3019 [nucl-th]].

\bibitem{Oishi:2016avj}
T.~Oishi, M.~Kortelainen and A.~Pastore,
Phys. Rev. C \textbf{96} (2017) no.4, 044327
doi:10.1103/PhysRevC.96.044327
[arXiv:1606.03111 [nucl-th]].

\bibitem{Wang:2021atf}
S.~M.~Wang and W.~Nazarewicz,
Phys. Rev. Lett. \textbf{126} (2021) no.14, 142501
doi:10.1103/PhysRevLett.126.142501
[arXiv:2104.03195 [nucl-th]].

\bibitem{Hagino:2008vm}
K.~Hagino, N.~Takahashi and H.~Sagawa,
Phys. Rev. C \textbf{77} (2008), 054317
doi:10.1103/PhysRevC.77.054317
[arXiv:0803.3258 [nucl-th]].

\bibitem{Garrido:2025qfv}
E.~Garrido and A.~S.~Jensen,
J. Phys. G \textbf{52} (2025) no.10, 105101
doi:10.1088/1361-6471/ae0959
[arXiv:2509.17908 [nucl-th]].

\bibitem{Kanada-Enyo:2007iri}
Y.~Kanada-En'yo,
Phys. Rev. C \textbf{76} (2007), 044323
[erratum: Phys. Rev. C \textbf{101} (2020) no.6, 069902]
doi:10.1103/PhysRevC.76.044323
[arXiv:0707.2120 [nucl-th]].

\bibitem{Aoyama:2009zz}
S.~Aoyama and N.~Itagaki,
Phys. Rev. C \textbf{80} (2009), 021304
doi:10.1103/PhysRevC.80.021304

\bibitem{Itagaki:2008zz}
N.~Itagaki, M.~Ito, K.~Arai, S.~Aoyama and T.~Kokalova,
Phys. Rev. C \textbf{78} (2008), 017306
doi:10.1103/PhysRevC.78.017306

\bibitem{Kobayashi:2013iwa}
F.~Kobayashi and Y.~Kanada-En'yo,
Phys. Rev. C \textbf{88} (2013) no.3, 034321
doi:10.1103/PhysRevC.88.034321
[arXiv:1305.4323 [nucl-th]].

\bibitem{Yamaguchi:2023xsx}
Y.~Yamaguchi, W.~Horiuchi, T.~Ichikawa and N.~Itagaki,
Phys. Rev. C \textbf{108} (2023) no.1, L011304
doi:10.1103/PhysRevC.108.L011304
[arXiv:2305.15624 [nucl-th]].

\bibitem{Nakagawa:2025zls}
K.~Nakagawa and Y.~Kanada-En'yo,
Phys. Rev. C \textbf{112} (2025) no.3, 034309
doi:10.1103/tjnl-jnxp
[arXiv:2505.16235 [nucl-th]].


\bibitem{Weiss:2023laq}
R.~Weiss and S.~Gandolfi,
Phys. Rev. C \textbf{108} (2023) no.2, L021301
doi:10.1103/PhysRevC.108.L021301
[arXiv:2301.09605 [nucl-th]].


\bibitem{JACOB:1966wce}
G.~Jacob and T.~a.~j.~Maris,
Rev. Mod. Phys. \textbf{38} (1966), 121-142
doi:10.1103/RevModPhys.38.121

\bibitem{Jacob:1973tu}
G.~Jacob and T.~A.~J.~Maris,
Rev. Mod. Phys. \textbf{45} (1973), 6-21
doi:10.1103/RevModPhys.45.6

\bibitem{Yoshida:2025ptaf119}
K.~Yoshida and J.~Tanaka,
Prog.\ Theor.\ Exp.\ Phys.\ \textbf{2025}, ptaf119 (2025)
doi:10.1093/ptep/ptaf119
[arXiv:2412.16649 [nucl-th]].



\bibitem{Faestermann:2022meh}
T.~Faestermann, A.~Bergmaier, R.~Gernh{\"a}user, D.~Koll and M.~Mahgoub,
Phys. Lett. B \textbf{824} (2022), 136799
doi:10.1016/j.physletb.2021.136799

\bibitem{RIBF-SHARAQ11:2024cvz}
K.~Miki \textit{et al.} [RIBF-SHARAQ11 and RCNP-E502],
Phys. Rev. Lett. \textbf{133} (2024) no.1, 012501
doi:10.1103/PhysRevLett.133.012501


\bibitem{Lazauskas:2022mvq}
R.~Lazauskas, E.~Hiyama and J.~Carbonell,
Phys. Rev. Lett. \textbf{130} (2023) no.10, 102501
doi:10.1103/PhysRevLett.130.102501
[arXiv:2207.07575 [nucl-th]].

\bibitem{Mazur:2024byg}
I.~A.~Mazur, M.~K.~Efimenko, A.~I.~Mazur, I.~J.~Shin, V.~A.~Kulikov, A.~M.~Shirokov and J.~P.~Vary,
Phys. Rev. C \textbf{110} (2024) no.1, 014004
doi:10.1103/PhysRevC.110.014004
[arXiv:2403.18232 [nucl-th]].










\bibitem{Lahde:2019npb}
T.~A.~L{\"a}hde and U.-G.~Mei{\ss}ner,
Lect. Notes Phys. \textbf{957} (2019), 1-396
Springer, 2019,
ISBN 978-3-030-14187-5, 978-3-030-14189-9
doi:10.1007/978-3-030-14189-9

\bibitem{Lu:2018bat}
B.~N.~Lu, N.~Li, S.~Elhatisari, D.~Lee, E.~Epelbaum and U.-G.~Mei{\ss}ner,
Phys. Lett. B \textbf{797} (2019), 134863
doi:10.1016/j.physletb.2019.134863
[arXiv:1812.10928 [nucl-th]].

\bibitem{Lu:2021tab}
B.~N.~Lu, N.~Li, S.~Elhatisari, Y.~Z.~Ma, D.~Lee and U.-G.~Mei{\ss}ner,
Phys. Rev. Lett. \textbf{128} (2022) no.24, 242501
doi:10.1103/PhysRevLett.128.242501
[arXiv:2111.14191 [nucl-th]].


\bibitem{Elhatisari:2022zrb}
S.~Elhatisari, L.~Bovermann, Y.~Z.~Ma, E.~Epelbaum, D.~Frame, F.~Hildenbrand, M.~Kim, Y.~Kim, H.~Krebs and T.~A.~L{\"a}hde, \textit{et al.}
Nature \textbf{630} (2024) no.8015, 59-63
doi:10.1038/s41586-024-07422-z
[arXiv:2210.17488 [nucl-th]].


\bibitem{Vernon:2010Galaxy}
I.~Vernon, M.~Goldstein and R.~G.~Bower,
Bayesian Anal. \textbf{5}, no.4, 619--669 (2010)
doi:10.1214/10-BA524.


\bibitem{Vernon:2014Galaxy}
I.~Vernon, M.~Goldstein and R.~Bower,
Statist. Sci. \textbf{29}, 81--90 (2014)
doi:10.1214/13-STS440.

\bibitem{Hu:2021trw}
B.~S.~Hu, W.~G.~Jiang, T.~Miyagi, Z.~H.~Sun, A.~Ekstr{\"o}m, C.~Forss{\'e}n, G.~Hagen, J.~D.~Holt, T.~Papenbrock and S.~R.~Stroberg, \textit{et al.}
Nature Phys. \textbf{18} (2022) no.10, 1196-1200
doi:10.1038/s41567-023-02324-9
[arXiv:2112.01125 [nucl-th]].


\bibitem{supp}%
\BibitemOpen
\bibinfo {note} {See Supplemental Material for details on the uncertainty quantification of chiral forces, evaluation of observables with pinhole configurations, intrinsic densities, schematic illustration of four-neutron correlations in $^{7}$H, extrapolation in Euclidean time, finite-volume effects, experimental information on $^{6,7}$H, two-neutron correlations in $^{6}$He and $^{3}$H, intrinsic densities and two-neutron correlations (including dineutron sizes) in $^{8}$He and $^{7}$H, opening angle $\theta$ and $\varphi$ dependence in four-neutron correlation of $^{7}$H, distribution of the dihedral angle in $^{7}$H, and internal angular and radial distributions from the four-neutron correlations in $^{7}$H.}\BibitemShut {Stop}%


\bibitem{Elhatisari:2017eno}
S.~Elhatisari, E.~Epelbaum, H.~Krebs, T.~A.~L{\"a}hde, D.~Lee, N.~Li, B.~n.~Lu, U.-G.~Mei{\ss}ner and G.~Rupak,
Phys. Rev. Lett. \textbf{119} (2017) no.22, 222505
doi:10.1103/PhysRevLett.119.222505
[arXiv:1702.05177 [nucl-th]].


\bibitem{Wiringa:2000gb}
R.~B.~Wiringa, S.~C.~Pieper, J.~Carlson and V.~R.~Pandharipande,
Phys. Rev. C \textbf{62} (2000), 014001
doi:10.1103/PhysRevC.62.014001
[arXiv:nucl-th/0002022 [nucl-th]].

\bibitem{Otsuka:2022bcf}
T.~Otsuka, T.~Abe, T.~Yoshida, Y.~Tsunoda, N.~Shimizu, N.~Itagaki, Y.~Utsuno, J.~Vary, P.~Maris and H.~Ueno,
Nature Commun. \textbf{13} (2022) no.1, 2234
doi:10.1038/s41467-022-29582-0

\bibitem{Shen:2022bak}
S.~H.~Shen, S.~Elhatisari, T.~A.~L{\"a}hde, D.~Lee, B.~N.~Lu and U.-G.~Mei{\ss}ner,
Nature Commun. \textbf{14} (2023) no.1, 2777
doi:10.1038/s41467-023-38391-y
[arXiv:2202.13596 [nucl-th]].

\bibitem{Shen:2024qzi}
S.~H.~Shen, S.~Elhatisari, D.~Lee, U.-G.~Mei{\ss}ner and Z.~X.~Ren,
Phys. Rev. Lett. \textbf{134} (2025) no.16, 162503
doi:10.1103/PhysRevLett.134.162503
[arXiv:2411.14935 [nucl-th]].



\bibitem{Bertsch:1991zz}
G.~F.~Bertsch and H.~Esbensen,
Annals Phys. \textbf{209} (1991), 327
doi:10.1016/0003-4916(91)90033-5


\bibitem{Aleksandrov:1984H6}
D.~V.~Aleksandrov, E.~A.~Ganza, Yu.~A.~Glukhov, B.~G.~Novatsky, A.~A.~Ogloblin, and D.~N.~Stepanov,
Sov.\ J.\ Nucl.\ Phys.\ \textbf{39}, 323 (1984)
[Yad.\ Fiz.\ \textbf{39}, 513 (1984)].

\bibitem{Belozyorov:1986H456}
A.~V.~Belozyorov \textit{et al.},
Nucl.\ Phys.\ A \textbf{460}, 352 (1986).

\bibitem{Gurov:2003pv}
Y.~B.~Gurov, D.~V.~Aleshkin, S.~V.~Lapushkin, P.~V.~Morokhov, A.~V.~Panin, V.~A.~Pechkurov, N.~O.~Poroshin, V.~G.~Sandukovsky, M.~V.~Telkushev and B.~A.~Chernyshev,
JETP Lett. \textbf{78} (2003), 183-187
doi:10.1134/1.1622028


\bibitem{Caamano:2008zz}
M.~Caamano, D.~Cortina-Gil, W.~Mittig, H.~Savajols, M.~Chartier, C.~E.~Demonchy, B.~Fernandez, M.~B.~G.~Hornillos, A.~Gillibert and B.~Jurado, \textit{et al.}
Phys. Rev. C \textbf{78} (2008), 044001
doi:10.1103/PhysRevC.78.044001



\bibitem{Korsheninnikov:2000He7unique}
A.~A.~Korsheninnikov \textit{et al.},
Phys.\ Scr.\ \textbf{T88}, 199 (2000).

\bibitem{Caamano:2007zz}
M.~Caama\~{n}o, D.~Cortina-Gil, W.~Mittig, H.~Savajols, M.~Chartier, C.~E.~Demonchy, B.~Fern\'andez, M.~B.~G\'omez Hornillos, A.~Gillibert, B.~Jurado, O.~Kiselev, R.~Lemmon, A.~Obertelli, F.~Rejmund, M.~Rejmund, P.~Roussel-Chomaz and R.~Wolski,
Phys.\ Rev.\ Lett.\ \textbf{99} (2007), 062502
doi:10.1103/PhysRevLett.99.062502




\bibitem{Nikolskii:2010zz}
E.~Y.~Nikolskii, A.~A.~Korsheninnikov, H.~Otsu, H.~Suzuki, K.~Yoneda, H.~Baba, K.~Yamada, Y.~Kondo, N.~Aoi and A.~S.~Denikin, \textit{et al.}
Phys. Rev. C \textbf{81} (2010), 064606
doi:10.1103/PhysRevC.81.064606

\bibitem{Caamano:2020}
M.~Caama\~{n}o, T.~Roger, A.~M.~Moro, G.~F.~Grinyer, J.~Pancin, \textit{et al.},
EPJ Web Conf.\ \textbf{232} (2020), 04002
doi:10.1051/epjconf/202023204002




\bibitem{Lin:2021xrc}
Y.~H.~Lin, H.~W.~Hammer and U.-G.~Mei{\ss}ner,
Phys. Rev. Lett. \textbf{128} (2022) no.5, 052002
doi:10.1103/PhysRevLett.128.052002
[arXiv:2109.12961 [hep-ph]].

\bibitem{Zhang:2024wfd}
S.~Zhang, S.~Elhatisari, U.-G.~Mei{\ss}ner and S.~H.~Shen,
Phys. Lett. B \textbf{869} (2025), 139839
doi:10.1016/j.physletb.2025.139839
[arXiv:2411.17462 [nucl-th]].


\bibitem{Johnson:1982yq}
R.~C.~Johnson,
Phys. Lett. B \textbf{114} (1982), 147-151
doi:10.1016/0370-2693(82)90134-4

\bibitem{Lu:2014xfa}
B.~N.~Lu, T.~A.~L{\"a}hde, D.~Lee and U.~G.~Mei{\ss}ner,
Phys. Rev. D \textbf{90} (2014) no.3, 034507
doi:10.1103/PhysRevD.90.034507
[arXiv:1403.8056 [nucl-th]].


\bibitem{Tilley:2004zz}
D.~R.~Tilley, J.~H.~Kelley, J.~L.~Godwin, D.~J.~Millener, J.~E.~Purcell, C.~G.~Sheu and H.~R.~Weller,
Nucl. Phys. A \textbf{745}, 155-362 (2004)
doi:10.1016/j.nuclphysa.2004.09.059

\bibitem{Yang:2023hyq}
Z.~H.~Yang, Y.~L.~Ye, B.~Zhou, H.~Baba, R.~J.~Chen, Y.~C.~Ge, B.~S.~Hu, H.~Hua, D.~X.~Jiang and M.~Kimura, \textit{et al.}
Phys. Rev. Lett. \textbf{131} (2023) no.24, 242501
doi:10.1103/PhysRevLett.131.242501


\bibitem{Keeley:2007bc}
N.~Keeley, F.~Skaza, V.~Lapoux, N.~Alamanos, F.~Auger, A.~Drouart, A.~Gillibert, L.~Nalpas, E.~C.~Pollacco and R.~Raabe, \textit{et al.}
Phys. Lett. B \textbf{646} (2007), 222-226
doi:10.1016/j.physletb.2007.01.035

\bibitem{Skaza:2007fus}
F.~Skaza, N.~Alamanos, F.~Auger, A.~Drouart, A.~Gillibert, V.~Lapoux, L.~Nalpas, E.~Pollacco, N.~Keeley and R.~Wolski, \textit{et al.}
Nucl. Phys. A \textbf{788} (2007), 260-265
doi:10.1016/j.nuclphysa.2007.01.093

\bibitem{Holl:2021bxg}
M.~Holl, R.~Kanungo, Z.~H.~Sun, G.~Hagen, J.~A.~Lay, A.~M.~Moro, P.~Navr{\'a}til, T.~Papenbrock, M.~Alcorta and D.~Connolly, \textit{et al.}
Phys. Lett. B \textbf{822} (2021), 136710
doi:10.1016/j.physletb.2021.136710
[arXiv:2110.01592 [nucl-ex]].

\bibitem{Fossez:2018gae}
K.~Fossez, J.~Rotureau and W.~Nazarewicz,
Phys. Rev. C \textbf{98} (2018) no.6, 061302
doi:10.1103/PhysRevC.98.061302
[arXiv:1806.02936 [nucl-th]].

\bibitem{Li:2021tyy}
H.~H.~Li, J.~G.~Li, N.~Michel and W.~Zuo,
Phys. Rev. C \textbf{104} (2021) no.6, L061306
doi:10.1103/PhysRevC.104.L061306
[arXiv:2305.05414 [nucl-th]].

\bibitem{Hiyama:2022gzv}
E.~Hiyama, R.~Lazauskas and J.~Carbonell,
Phys. Lett. B \textbf{833} (2022), 137367
doi:10.1016/j.physletb.2022.137367
[arXiv:2207.04634 [nucl-th]].

\bibitem{Assie:2010zz}
M.~Assie, J.~A.~Scarpaci, D.~Lacroix, J.~C.~Angelique, D.~Bazin, D.~Beaumel, Y.~Blumenfeld, W.~N.~Catford, M.~Chabot and A.~Chatterjee, \textit{et al.}
Mod. Phys. Lett. A \textbf{25} (2010), 1846-1849
doi:10.1142/S0217732310000460



\bibitem{Mueller:2007dhq}
P.~Mueller, I.~A.~Sulai, A.~C.~C.~Villari, J.~A.~Alcantara-Nunez, R.~Alves-Conde, K.~Bailey, G.~W.~F.~Drake, M.~Dubois, C.~Eleon and G.~Gaubert, \textit{et al.}
Phys. Rev. Lett. \textbf{99} (2007), 252501
doi:10.1103/PhysRevLett.99.252501
[arXiv:0801.0601 [nucl-ex]].

\bibitem{Zhukov:1994zz}
M.~V.~Zhukov, A.~A.~Korsheninnikov and M.~H.~Smedberg,
Phys. Rev. C \textbf{50} (1994), R1-R4
doi:10.1103/PhysRevC.50.R1

\bibitem{Mei:2012aop}
P.~Mei and P.~Van~Isacker,
Annals Phys. \textbf{327} (2012), 1182--1201
doi:10.1016/j.aop.2011.12.003.

\end{thebibliography}

\begin{thebibliography}{7}%
\makeatletter

\bibitem{Elhatisari:2022zrb_sm}
S.~Elhatisari, L.~Bovermann, Y.~Z.~Ma, E.~Epelbaum, D.~Frame, F.~Hildenbrand, M.~Kim, Y.~Kim, H.~Krebs and T.~A.~L{\"a}hde, \textit{et al.}
Nature \textbf{630}, no.8015, 59-63 (2024)
doi:10.1038/s41586-024-07422-z
[arXiv:2210.17488 [nucl-th]].


\bibitem{Vernon:2010Galaxy_sm}
I.~Vernon, M.~Goldstein and R.~G.~Bower,
Bayesian Anal. \textbf{5}, no.4, 619--669 (2010)
doi:10.1214/10-BA524.


\bibitem{Vernon:2014Galaxy_sm}
I.~Vernon, M.~Goldstein and R.~Bower,
Statist. Sci. \textbf{29}, 81--90 (2014)
doi:10.1214/13-STS440.

\bibitem{Hu:2021trw_sm}
B.~S.~Hu, W.~G.~Jiang, T.~Miyagi, Z.~H.~Sun, A.~Ekstr{\"o}m, C.~Forss{\'e}n, G.~Hagen, J.~D.~Holt, T.~Papenbrock and S.~R.~Stroberg, \textit{et al.}
Nature Phys. \textbf{18}, no.10, 1196-1200 (2022)
doi:10.1038/s41567-023-02324-9
[arXiv:2112.01125 [nucl-th]].


\bibitem{Lu:2018bat_sm}
B.~N.~Lu, N.~Li, S.~Elhatisari, D.~Lee, E.~Epelbaum and U.-G.~Mei{\ss}ner,
Phys. Lett. B \textbf{797} (2019), 134863
doi:10.1016/j.physletb.2019.134863
[arXiv:1812.10928 [nucl-th]].

\bibitem{Tanaka:2021oll_sm}
J.~Tanaka, Z.~H.~Yang, S.~Typel, S.~Adachi, S.~W.~Bai, P.~van Beek, D.~Beaumel, Y.~Fujikawa, J.~X.~Han and S.~Heil, \textit{et al.}
Science \textbf{371} (2021) no.6526, 260-264
doi:10.1126/science.abe4688


\bibitem{Caurier:2004gf_sm}
E.~Caurier, G.~Martinez-Pinedo, F.~Nowacki, A.~Poves and A.~P.~Zuker,
Rev. Mod. Phys. \textbf{77} (2005), 427-488
doi:10.1103/RevModPhys.77.427
[arXiv:nucl-th/0402046 [nucl-th]].

\bibitem{Hjorth-Jensen:1995zrg_sm}
M.~Hjorth-Jensen, T.~T.~S.~Kuo and E.~Osnes,
Phys. Rept. \textbf{261} (1995), 125-270
doi:10.1016/0370-1573(95)00012-6

\bibitem{Coraggio:2008in_sm}
L.~Coraggio, A.~Covello, A.~Gargano, N.~Itaco and T.~T.~S.~Kuo,
Prog. Part. Nucl. Phys. \textbf{62} (2009), 135-182
doi:10.1016/j.ppnp.2008.06.001
[arXiv:0809.2144 [nucl-th]].

\bibitem{Hagen:2013nca_sm}
G.~Hagen, T.~Papenbrock, M.~Hjorth-Jensen and D.~J.~Dean,
Rept. Prog. Phys. \textbf{77} (2014) no.9, 096302
doi:10.1088/0034-4885/77/9/096302
[arXiv:1312.7872 [nucl-th]].

\bibitem{Hergert:2015awm_sm}
H.~Hergert, S.~K.~Bogner, T.~D.~Morris, A.~Schwenk and K.~Tsukiyama,
Phys. Rept. \textbf{621} (2016), 165-222
doi:10.1016/j.physrep.2015.12.007
[arXiv:1512.06956 [nucl-th]].

\bibitem{Tohsaki:2001an_sm}
A.~Tohsaki, H.~Horiuchi, P.~Schuck and G.~Ropke,
Phys. Rev. Lett. \textbf{87} (2001), 192501
doi:10.1103/PhysRevLett.87.192501
[arXiv:nucl-th/0110014 [nucl-th]].

\bibitem{Funaki:2015uya_sm}
Y.~Funaki, H.~Horiuchi and A.~Tohsaki,
Prog. Part. Nucl. Phys. \textbf{82} (2015), 78-132
doi:10.1016/j.ppnp.2015.01.001

\bibitem{Zhou:2019cjz_sm}
B.~Zhou, Y.~Funaki, H.~Horiuchi and A.~Tohsaki,
Front. Phys. (Beijing) \textbf{15} (2020) no.1, 14401
doi:10.1007/s11467-019-0917-0
[arXiv:1905.00788 [nucl-th]].

\bibitem{Lahde:2014sla_sm}
T.~A.~L{\"a}hde, E.~Epelbaum, H.~Krebs, D.~Lee, U.-G.~Mei{\ss}ner and G.~Rupak,
J. Phys. G \textbf{42} (2015) no.3, 034012
doi:10.1088/0954-3899/42/3/034012
[arXiv:1409.7538 [nucl-th]].

\bibitem{He:2019ipt_sm}
R.~He, N.~Li, B.~N.~Lu and D.~Lee,
Phys. Rev. A \textbf{101} (2020) no.6, 063615
doi:10.1103/PhysRevA.101.063615
[arXiv:1910.01257 [cond-mat.quant-gas]].

\bibitem{Shen:2023_sm}
S.~H.~Shen, S.~Elhatisari, T.~A.~L{\"a}hde, D.~Lee, B.~N.~Lu and U.-G.~Mei{\ss}ner,
Nature Commun. \textbf{14} (2023) no.1, 2777
doi:10.1038/s41467-023-38391-y
[arXiv:2202.13596 [nucl-th]].

\bibitem{Aleksandrov:1984H6_sm}
D.~V.~Aleksandrov, E.~A.~Ganza, Yu.~A.~Glukhov, B.~G.~Novatsky, A.~A.~Ogloblin, and D.~N.~Stepanov,
Sov.\ J.\ Nucl.\ Phys.\ \textbf{39}, 323 (1984)
[Yad.\ Fiz.\ \textbf{39}, 513 (1984)].

\bibitem{Belozyorov:1986H456_sm}
A.~V.~Belozyorov \textit{et al.},
Nucl.\ Phys.\ A \textbf{460}, 352 (1986).

\bibitem{Gurov:2003pv_sm}
Y.~B.~Gurov, D.~V.~Aleshkin, S.~V.~Lapushkin, P.~V.~Morokhov, A.~V.~Panin, V.~A.~Pechkurov, N.~O.~Poroshin, V.~G.~Sandukovsky, M.~V.~Telkushev and B.~A.~Chernyshev,
JETP Lett. \textbf{78} (2003), 183-187
doi:10.1134/1.1622028


\bibitem{Caamano:2008zz_sm}
M.~Caamano, D.~Cortina-Gil, W.~Mittig, H.~Savajols, M.~Chartier, C.~E.~Demonchy, B.~Fernandez, M.~B.~G.~Hornillos, A.~Gillibert and B.~Jurado, \textit{et al.}
Phys. Rev. C \textbf{78} (2008), 044001
doi:10.1103/PhysRevC.78.044001

\bibitem{Nikolskii:2021kqe_sm}
E.~Y.~Nikolskii, I.~A.~Muzalevskii, A.~A.~Bezbakh, V.~Chudoba, S.~A.~Krupko, S.~G.~Belogurov, D.~Biare, A.~S.~Fomichev, E.~M.~Gazeeva and A.~V.~Gorshkov, \textit{et al.}
Phys. Rev. C \textbf{105} (2022) no.6, 064605
doi:10.1103/PhysRevC.105.064605
[arXiv:2105.04435 [nucl-ex]].

\bibitem{A1:2025mjf_sm}
T.~Shao \textit{et al.} [A1],
Phys. Rev. Lett. \textbf{134} (2025) no.16, 162501
doi:10.1103/PhysRevLett.134.162501
[arXiv:2501.01232 [nucl-ex]].

\bibitem{Korsheninnikov:2000He7unique_sm}
A.~A.~Korsheninnikov \textit{et al.},
Phys.\ Scr.\ \textbf{T88}, 199 (2000).




\bibitem{Golovkov:2004lsg_sm}
M.~S.~Golovkov, L.~V.~Grigorenko, A.~S.~Fomichev, Y.~T.~Oganessian, Y.~I.~Orlov, A.~M.~Rodin, S.~I.~Sidorchuk, R.~S.~Slepnev, S.~V.~Stepantsov and G.~M.~Ter-Akopian, \textit{et al.}
Phys. Lett. B \textbf{588} (2004), 163-171
doi:10.1016/j.physletb.2004.02.069

\bibitem{Caamano:2007zz_sm}
M.~Caama\~{n}o, D.~Cortina-Gil, W.~Mittig, H.~Savajols, M.~Chartier, C.~E.~Demonchy, B.~Fern\'andez, M.~B.~G\'omez Hornillos, A.~Gillibert, B.~Jurado, O.~Kiselev, R.~Lemmon, A.~Obertelli, F.~Rejmund, M.~Rejmund, P.~Roussel-Chomaz and R.~Wolski,
Phys.\ Rev.\ Lett.\ \textbf{99} (2007), 062502
doi:10.1103/PhysRevLett.99.062502




\bibitem{Nikolskii:2010zz_sm}
E.~Y.~Nikolskii, A.~A.~Korsheninnikov, H.~Otsu, H.~Suzuki, K.~Yoneda, H.~Baba, K.~Yamada, Y.~Kondo, N.~Aoi and A.~S.~Denikin, \textit{et al.}
Phys. Rev. C \textbf{81} (2010), 064606
doi:10.1103/PhysRevC.81.064606

\bibitem{Bezbakh:2019dvh_sm}
A.~A.~Bezbakh, V.~Chudoba, S.~A.~Krupko, S.~G.~Belogurov, D.~Biare, A.~S.~Fomichev, E.~M.~Gazeeva, A.~V.~Gorshkov, L.~V.~Grigorenko and G.~Kaminski, \textit{et al.}
Phys. Rev. Lett. \textbf{124} (2020) no.2, 022502
doi:10.1103/PhysRevLett.124.022502
[arXiv:1906.07818 [nucl-ex]].

\bibitem{Caamano:2020_sm}
M.~Caama\~{n}o, T.~Roger, A.~M.~Moro, G.~F.~Grinyer, J.~Pancin, \textit{et al.},
EPJ Web Conf.\ \textbf{232} (2020), 04002
doi:10.1051/epjconf/202023204002


\bibitem{Muzalevskii:2020svp_sm}
I.~A.~Muzalevskii, A.~A.~Bezbakh, E.~Y.~Nikolskii, V.~Chudoba, S.~A.~Krupko, S.~G.~Belogurov, D.~Biare, A.~S.~Fomichev, E.~M.~Gazeeva and A.~V.~Gorshkov, \textit{et al.}
Phys. Rev. C \textbf{103} (2021) no.4, 044313
doi:10.1103/PhysRevC.103.044313
[arXiv:2010.09655 [nucl-ex]].

\bibitem{Korsheninnikov:2003bz_sm}
A.~A.~Korsheninnikov, E.~Y.~Nikolskii, E.~A.~Kuzmin, A.~Ozawa, K.~Morimoto, F.~Tokanai, R.~Kanungo, I.~Tanihata, N.~K.~Timofeyuk and M.~S.~Golovkov, \textit{et al.}
Phys. Rev. Lett. \textbf{90} (2003), 082501
doi:10.1103/PhysRevLett.90.082501

\bibitem{Parker:1990plb_sm}
B.~Parker, K.~K.~Seth, and R.~Soundranayagam,
Phys.\ Lett.\ B \textbf{251}, 483 (1990),
doi:10.1016/0370-2693(90)90783-3.

\bibitem{Seth:1991zz_sm}
K.~K.~Seth and B.~Parker,
Phys. Rev. Lett. \textbf{66} (1991), 2448-2451
doi:10.1103/PhysRevLett.66.2448

\bibitem{Evseev:1981dce_sm}
V.~S.~Evseev, V.~S.~Kurbatov, V.~M.~Sidorov, 
V.~B.~Belyaev, J.~Wrzecionko, M.~Daum, R.~Frosch, 
J.~McCulloch, and E.~Steiner,
Nucl.\ Phys.\ A \textbf{352}, 379 (1981).


\bibitem{Aleksandrov:1982cf_sm}
D.~V.~Aleksandrov, Yu.~A.~Glukhov, A.~S.~Dem'yanova, 
et al.,
Sov.\ J.\ Nucl.\ Phys.\ \textbf{36}, 783 (1982).

\bibitem{Hagino:2005we_sm}
K.~Hagino and H.~Sagawa,
Phys. Rev. C \textbf{72} (2005), 044321
doi:10.1103/PhysRevC.72.044321
[arXiv:nucl-th/0508058 [nucl-th]].


\bibitem{Papadimitriou:2011jx_sm}
G.~Papadimitriou, A.~T.~Kruppa, N.~Michel, W.~Nazarewicz, M.~Ploszajczak and J.~Rotureau,
Phys. Rev. C \textbf{84} (2011), 051304
doi:10.1103/PhysRevC.84.051304
[arXiv:1109.0223 [nucl-th]].


\bibitem{Assie:2010zz_sm}
M.~Assie, J.~A.~Scarpaci, D.~Lacroix, J.~C.~Angelique, D.~Bazin, D.~Beaumel, Y.~Blumenfeld, W.~N.~Catford, M.~Chabot and A.~Chatterjee, \textit{et al.}
Mod. Phys. Lett. A \textbf{25} (2010), 1846-1849
doi:10.1142/S0217732310000460


\bibitem{Zhukov:1993aw_sm}
M.~V.~Zhukov, B.~V.~Danilin, D.~V.~Fedorov, J.~M.~Bang, I.~J.~Thompson and J.~S.~Vaagen,
Phys. Rept. \textbf{231} (1993), 151-199
doi:10.1016/0370-1573(93)90141-Y

\bibitem{Pillet:2007hb_sm}
N.~Pillet, N.~Sandulescu and P.~Schuck,
Phys. Rev. C \textbf{76} (2007), 024310
doi:10.1103/PhysRevC.76.024310
[arXiv:nucl-th/0701086 [nucl-th]].



\end{thebibliography}
\end{document}